\documentclass[usenatbib, twocolumn, a4paper,nofootinbib, linenumbers]{mnras}

\usepackage[utf8]{inputenc}  
\usepackage[T1]{fontenc}
\usepackage{graphicx,psfrag}
\usepackage{mathrsfs}
\usepackage{amsmath,amsfonts,amssymb,pifont,gensymb}
\usepackage{multirow,enumerate}
\usepackage{comment,hyperref}
\usepackage{color}
\usepackage{acronym}
\usepackage{xspace}
\usepackage[normalem]{ulem}
\usepackage{mathtools}
\usepackage{subfigure}
\usepackage{makecell}
\usepackage{changepage}
\usepackage{appendix}
\usepackage{lipsum}
\usepackage{natbib}

\newacro{BH}{black hole}
\newacro{NS}{neutron star}
\newacro{PN}{Post-Newtonian}
\newacro{BBH}{binary black hole}
\newacro{BNS}{binary neutron star}
\newacro{EOB}{effective-one-body}
\newacro{NR}{numerical relativity}
\newacro{GW}{gravitational wave}
\newacro{EOS}{equation-of-state}

\newcommand{\be}{\begin{equation}}
\newcommand{\ee}{\end{equation}}
\newcommand{\bea}{\begin{eqnarray}}
\newcommand{\eea}{\end{eqnarray}}
\newcommand{\bel}{\begin{align}}
\newcommand{\eel}{\end{align}}

\def\GMc2{{\rm G M_{\odot} c^{-2}}}

\def\SEOBNRv4T{\texttt{SEOBNRv4T}\xspace}

\usepackage{color}
\definecolor{cyan}{rgb}{0,0.9,0.9}
\definecolor{orange}{rgb}{0.9,0.5,0}
\definecolor{magenta}{rgb}{1,0,1}
\definecolor{purple}{rgb}{0.8,0.4,0.8}
\definecolor{gray}{rgb}{0.5,0.5,0.5}
\definecolor{mygreen}{rgb}{0.1,0.8,0.1}
\definecolor{darkblue}{rgb}{0.0,0.0,0.6}

\newcommand{\hypo}{\mathcal{H}_s}

\title[Parameter estimation methods for analyzing overlapping gravitational wave signals]{Parameter estimation methods for analyzing overlapping gravitational wave signals in the third-generation detector era}

\author[J. Janquart et al.]{Justin Janquart,$^{1, 2}$\thanks{E-mail: j.janquart@uu.nl}
Tomasz Baka$^{1, 2}$
Anuradha Samajdar$^{3}$
Tim Dietrich$^{3, 4}$
and Chris Van Den Broeck$^{1, 2}$
\\
$^{1}$Nikhef – National Institute for Subatomic Physics, Science Park, 1098 XG Amsterdam, The Netherlands\\
$^{2}$Institute for Gravitational and Subatomic Physics (GRASP), Department of Physics, Utrecht University,\\ Princetonplein 1, 3584 CC Utrecht, The Netherlands \\
$^{3}$Institut f\"{u}r Physik und Astronomie, Universit\"{a}t Potsdam, Haus 28, Karl-Liebknecht-Str. 24/25, 14476, Potsdam, Germany\\
$^{4}$Max Planck Institute for Gravitational Physics (Albert Einstein Institute), Am M\"uhlenberg 1, Potsdam, Germany
}

\date{\today}
\pubyear{2023}

\begin{document}

\maketitle

\begin{abstract}
\noindent
 In the coming years, third-generation detectors such as  Einstein Telescope and Cosmic Explorer will enter the network of ground-based gravitational-wave detectors. Their current design predicts a significantly improved sensitivity band with a lower minimum frequency than existing detectors. This, combined with the increased arm length, leads to two major effects: the detection of more signals and the detection of longer signals. Both will result in a large number of overlapping signals.
 It has been shown that such overlapping signals can lead to biases in the recovered parameters, which would adversely affect the science extracted from the observed binary merger signals. In this work, we analyze overlapping binary black hole coalescences with two methods to analyze multi-signal observations: \textit{hierarchical subtraction} and \textit{joint parameter estimation}. We find that these methods enable a reliable parameter extraction in most cases and that joint parameter estimation is usually more precise but comes with higher computational costs.
\end{abstract}


\section{Introduction}

The observation of gravitational waves (GWs) originating from compact binary coalescences (CBCs) is now done routinely with the advanced LIGO~\citep{Aasi_2015} and advanced Virgo~\citep{TheVirgo:2014hva} detectors, with tens of detections reported in the O3 GW catalog~\citep{LIGOScientific:2021djp}. The observations from this new information channel have had major impacts in fundamental physics~\citep{LIGOScientific:2021sio}, astrophysics~\citep{LIGOScientific:2021psn}, and cosmology~\citep{LIGOScientific:2021aug}. Moreover, the possibility to detect electromagnetic counterparts for binary neutron stars (BNSs), such as GW170817~\citep{LIGOScientific:2017ync}, has opened the perspective to do new multi-messenger studies. Further upgrades to the second generation detectors, as well as the addition of new detectors such as KAGRA~\citep{Somiya:2011np, Aso:2013eba, Akutsu:2018axf, Akutsu:2020his} and LIGO India~\citep{LigoIndia}, will lead to the observation of numerous events in the coming years. Moreover, going from these second-generation (2G) detectors to third-generation (3G) detectors, such as Einstein Telescope (ET)~\citep{Punturo_2010, Hild:2010id} and Cosmic Explorer (CE)~\citep{Reitze:2019iox, Abbott_2017, PhysRevLett.118.151105}, will lead to a major jump in the detection rate as well as in the duration of the signal due to the combined global increase in sensitivity and the major enhancements for lower frequencies~\citep{Sathyaprakash:2012jk}. In turn, this will lead to a high probability of CBC signals overlapping in the 3G detectors~\citep{Regimbau:2009rk, Samajdar:2021egv, Pizzati:2021apa, Relton:2021cax, Himemoto:2021ukb}. 

Previous works have studied the impact of such overlapping signals on data analysis when recovering one of the two signals and neglecting the presence of the other~\citep{Samajdar:2021egv, Pizzati:2021apa, Relton:2021cax, Himemoto:2021ukb, Antonelli:2021vwg}. These works employed different techniques but all have the same basic conclusions: bias can occur in various scenarios, and is most likely when the signals merge close to each other. 
In Ref.~\citep{Pizzati:2021apa}, using a Fisher matrix approach, the authors show that BNSs are less correlated in overlapping signals so that their bias becomes important only for close merger times ($< 0.1$ s), while the correlation between binary black holes (BBHs) is more important, meaning that the bias can happen for larger differences between the merger times. They then proceed to parameter estimation (PE) for overlapping BBHs in a LIGO-Virgo network varying some of the parameters, showing the appearance of biases for merger times close to each other. In~\citet{Himemoto:2021ukb}, the authors use Fisher matrices to study the bias that can occur in the parameters in both overlapping BBHs and BNSs, also finding that the bias becomes more important for short differences in the time of arrival. In~\citet{Relton:2021cax}, the authors focus on a LIGO Voyager scenario, looking at the biases not only based on their difference in merger times, but also as a function of other parameters, such as the sky location. For overlapping BBHs, they find a more important bias for closer merger times. However, they show that the observed bias for a given difference in merger time can change substantially depending on the sky location of the two events. Moreover, two overlapping BBHs can be mistaken for one strongly precessing BBH. Furthermore, the authors suggest that for overlapping BNS and BBH signals no major bias will occur due to the different durations of the two signals. This is corroborated by the analyses done in~\citet{Samajdar:2021egv}, where three scenarios are analyzed: two overlapping BBH signals, two overlapping BNS signals and the overlap of a BBH of varying masses with a BNS. In the latter case, the authors find that there is hardly any effect on the BNS parameter estimation, probably due to the difference in the number of cycles present in-band for this signal. However, in this scenario, the BBH can be affected by significant bias, especially when the BBH has high component masses. The bias mostly disappears when the merger times are separated by more than two seconds. For two overlapping BBHs, if the total masses and hence the durations are different, the PE is done relatively well. However, others~\citep{Pizzati:2021apa, Himemoto:2021ukb, Relton:2021cax} have shown that, if the two BBHs have similar source properties, biases can be present. Finally, for two overlapping BNSs, it appears that the signal with the highest SNR is relatively well recovered in all cases, but not necessarily the quieter signal. 

One problem that is not covered by previous studies is the effect of confusion noise on PE. Indeed, the high rate of the events and their duration will make for very few periods without signal in-band for the 3G detectors~\citep{Samajdar:2021egv}. As a consequence, it will be very difficult to estimate the noise present in the detectors, and additional biases can occur due to a mismodelling of the noise~\citep{Antonelli:2021vwg}\footnote{We will not consider this in this work as we will only look at overlapping binary signals.}. A demonstration of the effect of this noise on matched filtering and how the PSD could be computed are presented in~\citet{Wu:2022pyg}.

The presence of biases when signals merge close to each other and the relative occurrence of such scenarios based on the estimated rates shows that PE methods will have to be adapted to be suited for the 3G cases. Indeed, biases in the parameters estimated for the CBCs would impact any direct science case for the CBCs (such as measuring their mass distribution and rate, or testing general relativity~\citep{Hu:2022bji}) and also some indirectly related ones, such as the search for primordial backgrounds since this requires the subtraction of the foreground sources~\citep{Sachdev2020:pol, Sharma:2022kds, Biscoveanu:2020ste, Zhou:2022nmt, Zhou:2022otw, Reali:2022aps}.

In this work, we look at two possible methods to analyze overlapping signals. One is \emph{hierarchical subtraction} (HS), where we analyze one signal (typically the loudest), then subtract the maximum-likelihood template before analyzing the second one. However, if an important bias happens when analyzing the first signal, the parameters of both events might be biased\footnote{In this work, the term ``bias'' is used colloquially and denotes any changes in the recovered posteriors due to the presence of another signal.}. Optionally, one can also perform a third run, subtracting the maximum-likelihood parameters for the second event and re-analyzing the dominant signal to reduce the bias in its recovery. Still, this is not guaranteed to lead to unbiased results. Therefore, we also implement a \emph{joint parameter estimation} (JPE) framework, where the two signals are analyzed at the same time to account for the entire model. In principle, this should be the most complete model one can use. Due to the high dimensionality of the parameter space, combined with the large duration of the signals in the 3G detectors, this framework is substantially slower than HS\footnote{For our experiments, the two frameworks were run on the same cluster using 16 \emph{Intel(R) Xeon(R) Gold 6152} CPUs. The average run time for JPE is 23.8 days, while for HS, the first run took an average of 6.3 days, the second run an average of 4.3 days, and the last run took an average run time of 6.1 days. So, on average, JPE takes 7 more days to complete than HS if we perform the 3 runs. If one is satisfied with the two first runs, the difference between the two approaches goes up to about 2 weeks.}. It would be close to impossible to follow the predicted rates for an ET and CE network using a simple JPE like the one used in this work. Such constraints could be alleviated by using recently developed techniques, like relative binning~\citep{Zackay:2018qdy, Dai:2018dca, Leslie:2021ssu}, adaptive frequency resolution~\citep{Morisaki:2021ngj}, or machine learning~\citep{Dax:2021tsq, williams2021nested, Langendorff:2022fzq}. One could also count on the development of more powerful computational methods, such as quantum computing~\citep{Gao:2021rxg} but it is difficult to have an idea of the state of such methods by the time the 3G detectors get online. Still, it is important to start preparing for the future of 3G detectors now, hence, it is important to start looking at the parameter estimation of overlapping signals to have the bases to build upon. Due to the limited computational resources, this work focuses on the parameter estimation of two overlapping BBH signals.

This article is structured as follows. In Sec.~\ref{sec:methods}, we explain the two methods applied to perform the analysis of the overlapping signals. In the next section, we explain the setup of the analyses, while in Sec.~\ref{sec:results} we present the results of the analyses. Finally, Sec.~\ref{sec:conclusions} provides conclusions and outlook.

\section{Description of the methods}
\label{sec:methods}

When performing GW data analysis on CBC signals, our objective is to find the posterior probability density function (PDF) of the binaries' parameters ($\theta$): $p(\theta | d, \hypo)$, where $d$ represents the data, and $\hypo$ is the hypothesis under which we work (e.g.\ there is a GW signal in the data). Using Bayes' theorem~\citep{Veitch:2009hd}:
\begin{equation}\label{eq:BayesTheorem}
    p(\theta | d, \hypo) = \frac{p(d | \theta, \hypo)p(\theta | \hypo)}{p(d | \hypo)} \, ,
\end{equation}
where $p(\theta | \hypo)$ is the prior on the binary parameters, and $p(d | \theta, \hypo)$ is the likelihood for the data given a set of parameters. Finally, $p(d | \hypo)$ is called the evidence and represents a normalization factor for the posterior probability density:
\begin{equation}\label{eq:evidence}
    p(d | \hypo) = \int d\theta p(d | \theta, \hypo) p(\theta | \hypo) \, .
\end{equation}

For GW inference, $d(t)$ is the output of the interferometers, which can be seen as made of a noise component $n(t)$ and, under the signal hypothesis, a GW component $h(t)$:
\begin{equation}\label{eq:strain}
    d(t) = n(t) + h(t) \,.
\end{equation}
In our scenario, the GW component can consist of one or more signals. In the latter case, $h(t) = \sum_{i = 1}^{N} h_{i}(t)$, where $h_i(t)$ is the representation of each individual GW signal, and $N$ is the total number of GW signals present in the data stretch under consideration.

Assuming the noise to be Gaussian, the likelihood of having the data $d(t)$ given the presence of a GW component $h(t)$, is given by the proportionality
\begin{equation}\label{eq:SingleEventLikelihood}
    p(d | \theta, \hypo) \propto \exp\bigg[ -\frac{1}{2} \langle d - h(\theta) | d - h(\theta) \rangle \bigg] \, ,
\end{equation}
where
\begin{equation}\label{eq:InnerProduct}
    \langle a | b \rangle = 4 \Re \int_{f_{\rm{min}}}^{f_{\rm{max}}}  \frac{\tilde{a}^{*}(f)b(f)}{S_{n}(f)} df \,.
\end{equation}
In this expression, $\tilde{a}(f)$ refers to the Fourier transform of the function $a(t)$, $^{*}$ is the complex conjugate, and $S_n(f)$ is the power spectral density (PSD). $f_{\rm{min}}$ and $f_{\rm{max}}$ represent respectively the lower and the upper frequency chosen for the analysis\footnote{Usually, $f_{\rm{min}}$ is chosen to be the lowest frequency at which the noise is approximately stationary, and $f_{\rm{max}}$ is the Nyquist frequency.}. 

\subsection{Joint Parameter estimation}
When the noise component in Eq.~\eqref{eq:strain} is made of multiple signals, the likelihood in Eq.~\eqref{eq:SingleEventLikelihood} becomes
\begin{align}\label{eq:JointLikeli}
     & p(d | \theta, \hypo) \nonumber\\ & \propto \exp \Bigg[ -\frac{1}{2} \bigg\langle d - \sum_{i = 1}^{N} h(\theta_i) \bigg| d - \sum_{i = 1}^{N} h(\theta_i)\bigg\rangle  \Bigg] \, ,
\end{align}
where we just expanded the expression for $h(t)$ compared to the previous expression. Here, $N$ is the total number of signals, and $\theta_i$ represents the set of parameters describing the $i^{th}$ GW signal so that $\theta = \{ \theta_1, \dots, \theta_i , \dots, \theta_N \}$ in this case. 

Using Eq.~\eqref{eq:JointLikeli} gives rise to the method of joint parameter estimation (JPE), where we jointly look for $\{\theta_i\}_{i = 1, \dots, N}$. Since each $\theta_i$ is a set of 15 parameters for spinning BBHs, a set of 16 parameters for NSBHs, and a set of 17 parameters for BNSs\footnote{Typically, a BBH is described by two mass parameters, 6 spin parameters, a distance parameter, the inclination, 2 parameters for the sky location, the merger time, the phase of coalescence, and the polarization angle. Usually, for each neutron star present in the system, one also adds a tidal deformability, however, for BNSs the dimensionality could increase even further if higher-order tidal contributions, spin-induced quadrupole effects, or resonant effects are also taken into account. 

In addition, formally, when parameter estimation is performed, one can also add the noise-related calibration parameters, which would further increase the dimensionality of the parameter space.}, it means that, if we have $X$ BBHs, $Y$ NSBHs, and $Z$ BNSs (so $N = X + Y + Z$), the parameter space has $15 X + 16 Y + 17 Z$ parameters to explore. In the end, this means that as soon as we consider 2 signals, the parameter space grows to at least 30 dimensions, which is already challenging with our traditional methods, showing the difficulty to analyze several signals jointly.

In addition, we can assume that there is some uncertainty on the total number of signals $N$ in the data. In this case, it is also possible to sample over $N$ and the signal types, effectively allowing for any number of signals to be present in a given data stretch. 

The problem of joint analysis of several signals has already been looked at in other contexts, such as the characterization of the nearly monochromatic signals coming from white dwarf binaries in LISA~\citep{Littenberg:2020bxy}, BNSs in the Big Bang Observer~\citep{Cutler:2006plj}, or supermassive black holes in pulsar timing array searches~\citep{Petiteau:2013die}. However, the different characteristics of the signal looked for in these various context makes the methods different from one case to the other.

In this work, we will only consider the possibility to have two signals in the data. So, we write the data as
\begin{equation}\label{eq:TwoSignalsData}
    d(A, B, t) = h_{A}(t) + h_B(t) + n(t) \, ,
\end{equation}
where we just denote the signals by $A$ and $B$, without any importance on which signal is $A$ and which signal is $B$.

In this case, the likelihood takes the particular form
\begin{align}\label{eq:Likelihood2signals}
    & p(d(A, B) | \theta, \hypo)\nonumber \\ 
    &\propto \exp \Bigg[ -\frac{1}{2} \bigg\langle d(A, B) - h_A(\theta_A) - h_B(\theta_B) \bigg| \nonumber \\ 
    & d(A, B) - h_A(\theta_A) - h_B(\theta_B) \bigg\rangle  \Bigg] \, .
\end{align}
This is just the explicit form of Eq.~(\ref{eq:JointLikeli}) for two signals. 

In principle, if the sources are of the same nature, the labels $A$ and $B$ are interchangeable during the sampling, making the likelihood symmetric in two events.
In our algorithm, we do not impose any conditions on the parameters to break this symmetry. As a consequence, this is something that needs to be done in a post-processing step, as we need to assign drawn samples to the correct event. Sometimes, not accounting for this condition leads to bimodalities. In this work, we use a time ordering condition, taking the samples for event $A$ to be those that arrive first in time, and the samples for event $B$ to be those arriving second in time. In future work, this condition could directly be imposed in the algorithm by having a conditional prior such that the time of arrival of one event is always smaller than the other. We also note that the condition can be imposed on other parameters, such as the chirp mass for example.

\subsection{Hierarchical subtraction}
\label{sec:method_hier}
In hierarchical subtraction (HS), the idea is not to fit the two signals at once but to rather do a combination of individual signal analyses and subtraction of best-fit parameters. Therefore, we start by running a single event parameter estimation analysis on the data $d(A, B)$ to get the characteristics of the dominant signal. If we label by $A$ the loudest signal, we can denote the best-fit parameters (typically the maximum likelihood parameters) $\hat{\theta}_A$, and the waveform corresponding to this signal 
\begin{equation}\label{eq:BestFitA}
    \hat{h}_A(t) = h(t, \hat{ \theta}_A).
\end{equation}
Using this, we can get the data for signal B given signal A by subtracting the best-fit template
\begin{equation}\label{eq:SubtractedData}
    d(B, r_A, t) =  d(A, B, t) - \hat{h}_A(t) \, ,
\end{equation}
where $r_A$ are the residuals of signal A due to the imperfect subtraction. This is an approximate data strain for the second event in the data since the maximum likelihood parameters used to model the first event are prone to errors, with errors coming from the modeling itself but also from the presence of the second event when characterizing the first one. 

We can then analyze $d(B, r_A, t)$ to get the parameters for event $B$, leading to a posterior distribution for the two events. In principle, if the bias on the first recovery is not too important, then the posteriors on the second event should be fine. However, this approach is less robust than JPE, where we correctly account for the presence of several events. 

In Ref.~\citet{Antonelli:2021vwg}, the authors also suggest a way to correct the bias due to the individual characterization of two signals\footnote{In their paper, they also account for the possible confusion background due to the sum of all the mergers going on in the background.}. Once the two signals have been analyzed separately, we can use the two best-fitting posteriors to evaluate the bias made in the model reconstruction for each signal. The estimated biases can then be applied as a correction factor to the best-fitting parameters. We could then redo the subtraction of each event and analyze it again, but now with a subtracted signal that is closer to the real one, reducing the possible bias in the recovered posterior. Though this method is attractive, it requires multiple parameter estimation runs, which are expensive in a 3G detector context, and the computation of the biases also requires solving a combination of Fisher matrices and numerical derivatives, making it a non-trivial operation. 

\section{Setup of the analyses}
\label{sec:setups}
 
 \begin{table*}
    \centering
    \begin{tabular}{c c c}
    \hline
    \hline
    Parameter & Population generation & Prior \\
    \hline 
    \hline
    & & \\
    Primary component mass & PowerLaw + Peak~\citep{LIGOScientific:2021djp} & / \\
    Mass ratio &  PowerLaw + Peak~\citep{LIGOScientific:2021djp} & $\mathcal{U}(0.1, 1)$ \\
    Chirp mass & / & $\mathcal{U}(4, 200)$ \\
    Redshift & Oguri's fit~\citep{Oguri:2018muv} + rescaling & / \\
    Luminosity distance & / & Uniform comoving volume [$1$, $100$] Gpc \\
    Spin amplitude 1 & $\mathcal{U}(0, 1)$ & $\mathcal{U}(0, 1)$ \\
    Spin amplitude 2 & $\mathcal{U}(0, 1)$ & $\mathcal{U}(0, 1)$ \\
    Tilt angle 1 & Uniform in sine & Uniform in sine \\
    Tilt angle 2 & Uniform in sine & Uniform in sine \\
    Spin vector azimuthal angle & $\mathcal{U}(0, 2\pi)$ & $\mathcal{U}(0, 2\pi)$\\
    Precession angle about angular momentum & $\mathcal{U}(0, 2\pi)$ & $\mathcal{U}(0, 2\pi)$ \\
    Inclination angle & Uniform in sine & Uniform in sine \\
    Wave polarization & $\mathcal{U}(0, \pi)$ & $\mathcal{U}(0, \pi)$ \\
    Phase of coalescence & $\mathcal{U}(0, 2\pi)$ & $\mathcal{U}(0, 2\pi)$ \\
    Right ascension & $\mathcal{U}(0, 2\pi)$ & $\mathcal{U}(0, 2\pi)$ \\
    Declination & Uniform in cosine & Uniform in cosine \\
    Time of coalescence & Uniform over a year (second precision) & $\mathcal{U}(t_{inj} - 0.1, t_{inj} + 0.1)$ \\
     & & \\
     \hline
    \hline
    
    \end{tabular}
    \caption{Overview of the functions used to generate the different parameters for the BBH population and the priors used for the parameter estimation recovery.}
    \label{tab:params}
\end{table*}
 
\begin{figure}
     \centering
     \includegraphics[keepaspectratio, width=0.49\textwidth]{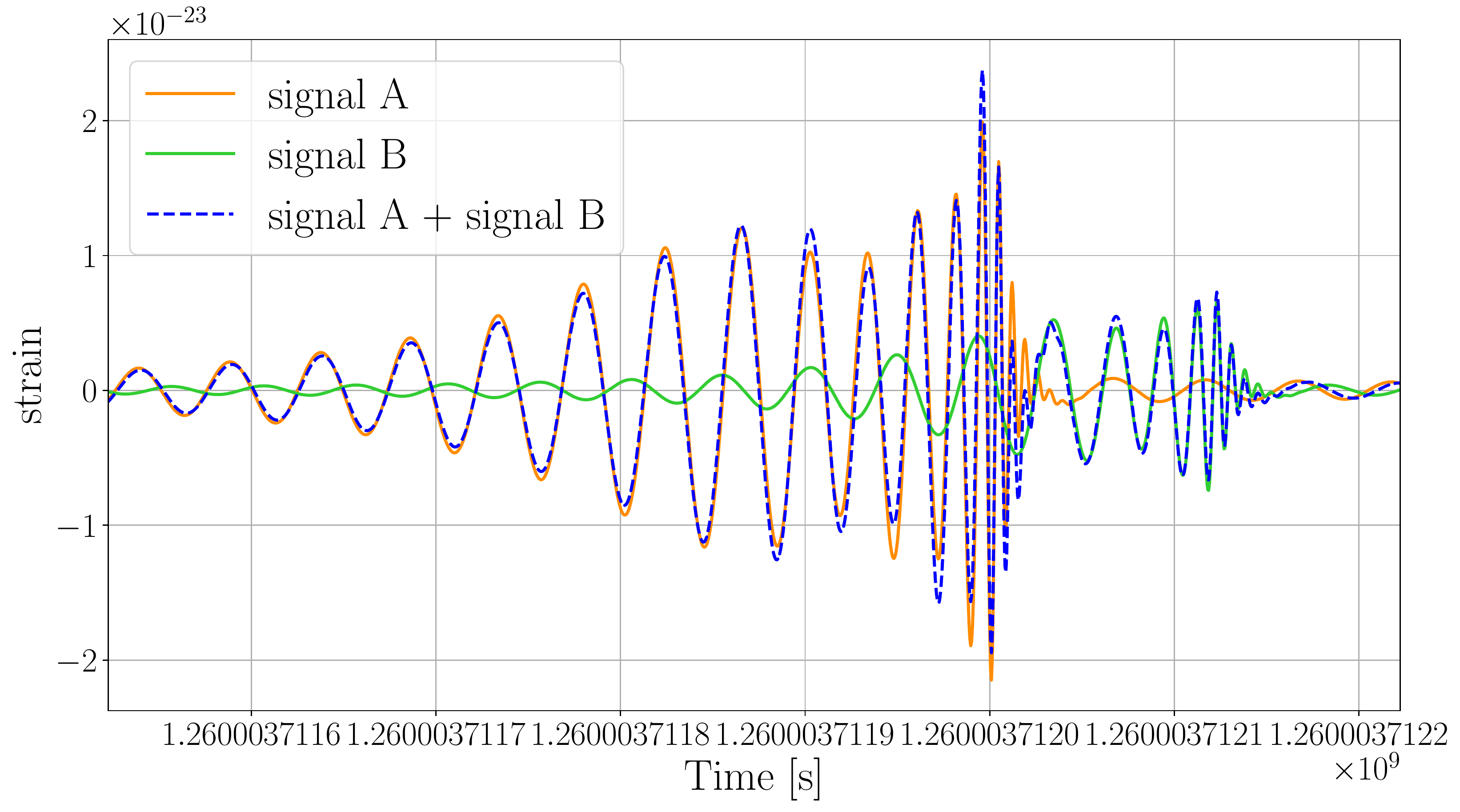}
     \includegraphics[keepaspectratio, width=0.49\textwidth]{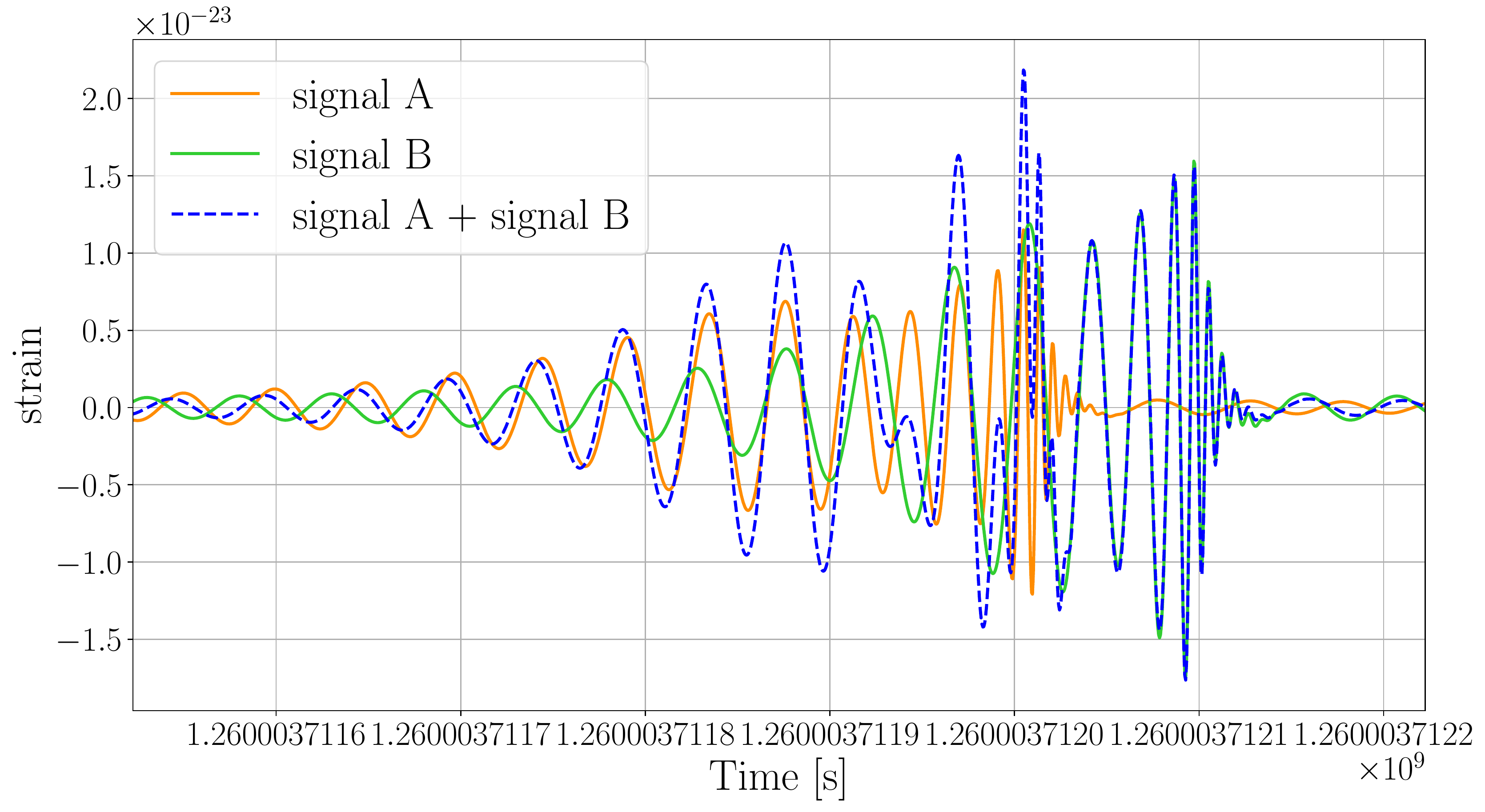}
     \caption{\textit{Top:} Representation of two overlapping BBHs and the underlying signals for signals with different SNRs ($\rm{SNR}_{\rm{A}}  = 46.1$ and $\rm{SNR}_{\rm{B}}  = 22.2$). In this case, the sum of the two signals is mostly dominated by the loudest signal, while the effect of the quieter signal takes up only after the loudest signal has merged. \textit{Bottom:} Representation of two overlapping BBHs and the underlying signals for signals with similar SNRs ($\rm{SNR}_{\rm{A}}  = 34.3$ and $\rm{SNR}_{\rm{B}}  = 30.2$). In this scenario, there is no signal dominating over the other for the entire duration of the event and the overlapping signals have more complicated features.}
     \label{fig:overlappedSignals}
 \end{figure}

 Due to the computational resources that would be required to analyze 3G signals, we focus on overlapping BBHs with masses in $[30, 60]M_{\odot}$\footnote{The higher masses are chosen to not have a signal with a too long duration while still enabling overlap for the difference in times of arrival used in this work.}. We use a network of detectors made of one triangular Einstein Telescope with 10 km arm-lengths, and a Cosmic explorer detector located at the LIGO-Hanford position and with 40km arm-lengths. We generate stationary Gaussian noise from the detectors' PSDs, where, for ET, we use the ET-D PSD~\citep{Punturo_2010, Hild:2010id}, and for CE, we use the projected PSD from Refs.~\citet{Abbott_2017, Reitze:2019iox}. We then inject two simulated BBH signals into the artificial noise. For this study, we take a lower cutoff on the signals of $20$ Hz. A representation of the waveforms obtained by the addition of two BBHs can be found in Fig.~\ref{fig:overlappedSignals}. One sees that the final signal has a non-trivial shape, illustrating the risk of biased posteriors when not accounting for the presence of two signals. In addition, one also sees that depending on the relative SNR of the signals, the observed deformation in the waveform is different. For the case where one signal has a significantly higher SNR than the other ($\sim \times 2$, top panel), the quieter signal will somewhat deform the signal, but the observed waveform will resemble mostly the loudest signal. On the other hand, for signals with close SNRs (bottom panel), we see that the deformation of the signal can be more complicated, without really having a dominant signal (except for the fraction of a second where one signal has merged and the other is still merging). Based on these observations, one can expect HS to be more effective when one signal clearly dominates over the other. 
 
In our study, we simulate 55 such mergers. To produce these high-mass signals, we sample the component masses from the Power-law + Peak distribution from~\citet{LIGOScientific:2021djp} but keep only the systems that fulfill the mass requirement. In addition, the events are sampled in redshift according to the merger rate density reconstructed from Oguri's fit~\citep{Oguri:2018muv}. The sky location is drawn to be uniform on the sky, and the spin parameters are picked from an isotropic distribution. For overlapping signal events, the coalescence time of the first event is drawn from a uniform distribution spanning over an entire year, while the second event is placed 0.1 seconds later. An overview of the functions used to make the binaries and the priors used for the analysis is given in Table~\ref{tab:params}.

Since the SNR of the signals can reach hundreds to thousands in an ET and CE network, and such high values make the computation time even longer, we decided to rescale the SNR to take values constrained between $8$ and $50$. This is done by adjusting the luminosity distance. However, since we expect the SNR ratio between the events to play a role, we try to keep this ratio as close as possible to the original one. So, if the loudest signal has a value above 50, we rescale it to take a value between 45 and 50 (this value is drawn randomly from a uniform distribution). Then, we rescale the quieter signal with the same factor. If this value is below 8, then we choose a new scaling factor to bring the SNR back between 8 and 13 (once more using a uniform distribution). Each system is then analyzed once without additional noise, and once injected in Gaussian noise generated from the PSDs.
 
 For the different runs, we choose fixed priors for the various parameters. The right column of Table.~\ref{tab:params} gives an overview of the priors used for the different parameters. In particular, we take a uniform prior on chirp mass ($\mathcal{M}_c = \frac{(m_1 m_2)^{3/5}}{(m_1 + m_2)^{1/5}}$) and mass ratio ($q = \frac{m_2}{m_1}$), with bounds of $[4, 200]M_\odot$, and $[0.1, 1]$. We also take a uniform in comoving volume prior for the luminosity distances, with bounds going from $1$ to 100 Gpc. These priors are adapted to cover any possible signal present in our set of data. The priors for the other parameters correspond to the usual priors taken for BBHs. When doing the JPE and HS runs, the priors are the same for the two events, and we do not add any conditions related to the signals (for example time ordering of the signals or enforcing one signal to be heavier than the other). 
 
We note here that an alternative approach is to use narrower priors informed by the results of low-latency searches~\citep{Regimbau:2012ir, Relton:2022whr}, which very likely could only be applied for a couple of parameters such as the chirp mass and the coalescence time. In~\citet{Relton:2022whr}, the authors show that matched filtering pipelines and unmodelled searches can pick up overlapping signals with reasonable accuracy. They also suggest some enhancements to make the pipelines even more suited for the challenge of overlapping signals detection. In addition, they show that for most of the overlapping signals, the error on the chirp mass is not much larger than for the non-overlapping case, even if for the occurrences where the signals are very close in time, the error increases. Therefore, using such searches to set narrower priors is a realistic alternative. However, currently, they also seem to contain risks, as an increased difference in the value recovered for some parameters can happen, and taking too narrow a prior could lead to the exclusion of the actual value from the prior. In the end, extra developments would be needed to make sure that using these results to narrow down the initial priors is viable.
 
To have a basis of comparison, we also do the parameter estimation of the individual signals. This is done by using the same priors as the one explained above but injecting only one of the two signals in the noise. 
 
 All the parameter estimation runs are performed using \textsc{bilby}~\citep{Ashton:2018jfp} with the \textsc{dynesty}~\citep{Speagle:2020} sampler. For the JPE runs, we added our own adapted joint likelihood in the package to keep a consistent framework. 

\section{Results and discussions}
\label{sec:results}
In this section, we show the results of the different approaches. We first compare the HS approach with single parameter estimation (SPE). In the latter, we only inject one of the two signals and perform PE on it. Then, we compare JPE with SPE and HS. Here, we focus on the results of the analyses performed with noise. The conclusion in the no-noise case are similar and can be found in Appendix~\ref{sec:AppNoNoise}. For all the figures presented in this work, when plotting individual event results, we represent by a dot and label as ``loud'' the loudest events in the pair, while we represent by a triangle and label as ``quiet'' the quieter ones.

\subsection{Hierachical subtraction}

\begin{figure*}
    \centering
    \includegraphics[keepaspectratio, width=0.49\textwidth]{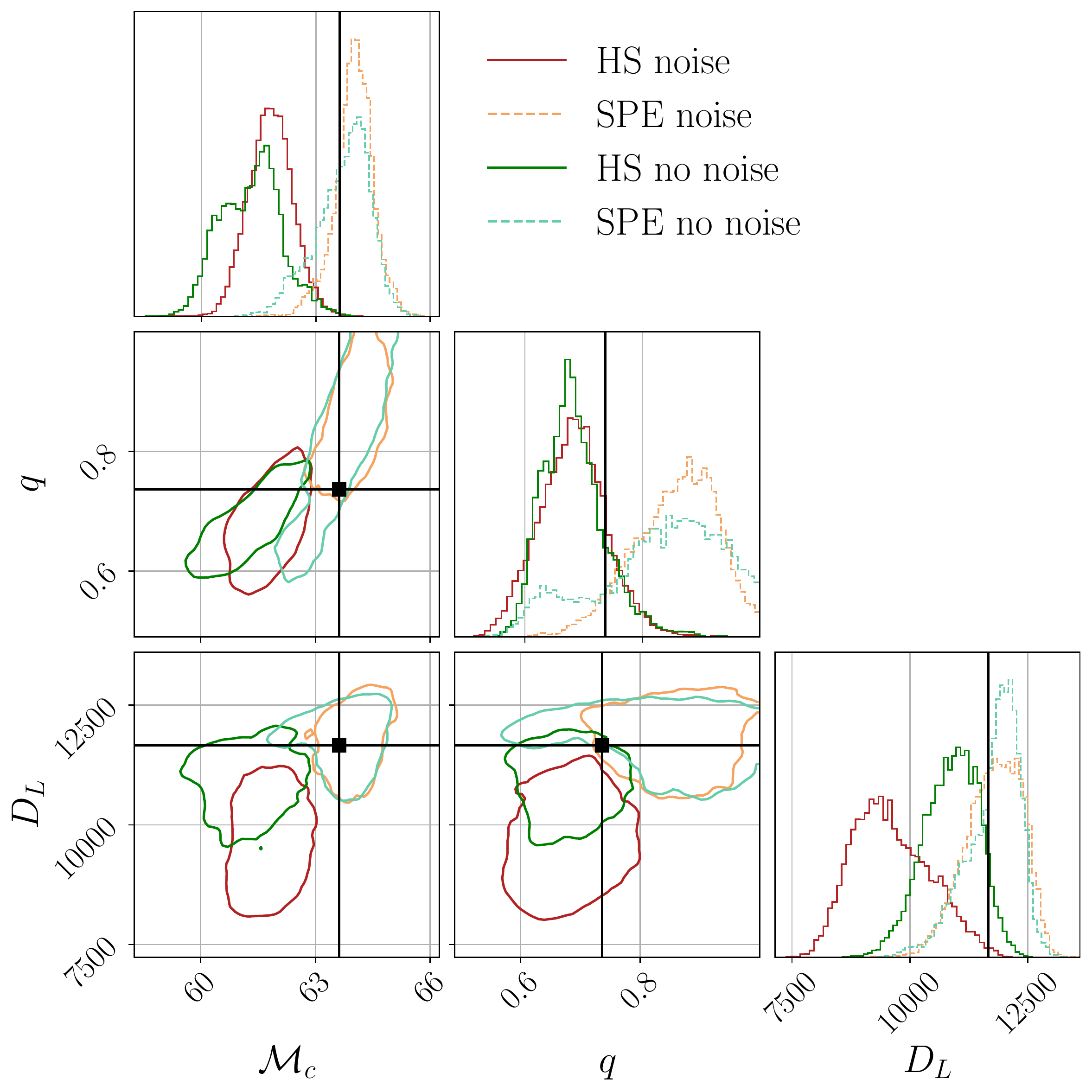}
    \includegraphics[keepaspectratio, width=0.49\textwidth]{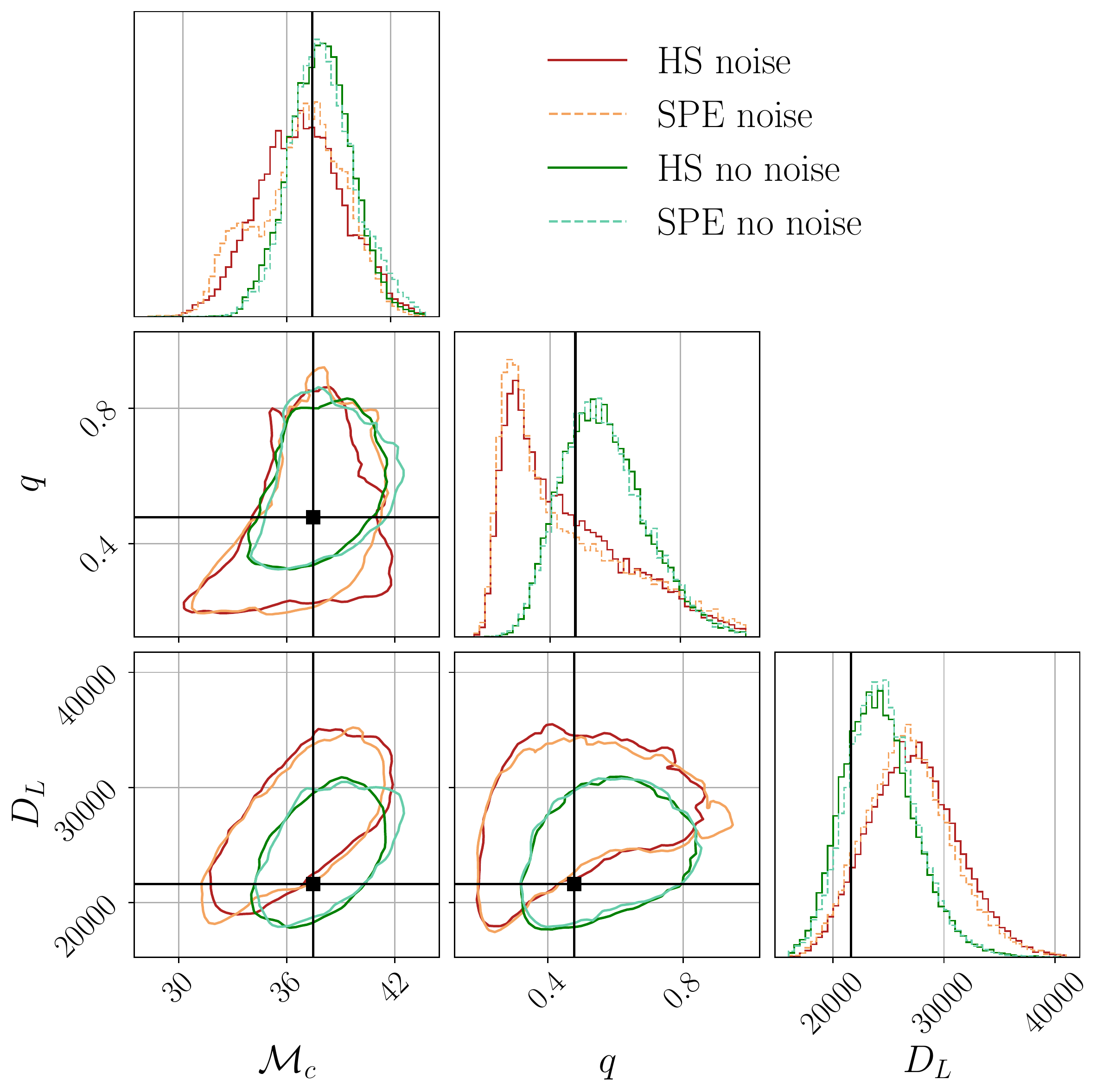}
    \caption{Comparison of the posteriors for the HS and the SPE methods with and without noise. We represent the chirp mass ($\mathcal{M}_c$), the mass ratio ($q$), and the luminosity distance ($D_L$). \emph{Left:} Case where HS is worse than SPE. HS posteriors are significantly shifted compared both to the injected value and SPE posteriors. \emph{Right:} Case where HS is close to SPE, with equivalent posteriors recovered in the two cases.}
    \label{fig:hierPosterior}
\end{figure*}

We start with discussing the result of the analysis for the HS approach. In almost all of the HS runs, the first PE stage picked up the signal with higher SNR. Nevertheless, there were two cases where it instead picked up the quieter signal; these were instances where the SNRs of the injected signals were close to each other. In such a case, the signal picked first is not the same with and without noise. A representative example of the posteriors can be found in Fig.~\ref{fig:hierPosterior}. While the widths of the distributions closely match the ones obtained with the SPE runs, in most cases they show bias in the recovered parameters.

When comparing the HS and the SPE recovery of the parameters, one sees that the HS recovery is nearly always biased. However, this bias seems to be more pronounced when there is also noise in the injection. For example, in Fig.~\ref{fig:hierPosterior}, one sees that the zero-noise HS recovery is close to the one for SPE with noise, while the recovery with noise is off. This shows that the mismodelling of the noise (due to an additional event) is present, as one could expect from previous works on biases in overlapping signals~\citep{Samajdar:2021egv, Relton:2021cax, Pizzati:2021apa, Antonelli:2021vwg}. However, this does not necessarily mean that all parameters are off, and the first signal's characteristics can be recovered.

\begin{figure}
    \centering
    \includegraphics[keepaspectratio, width=0.49\textwidth]{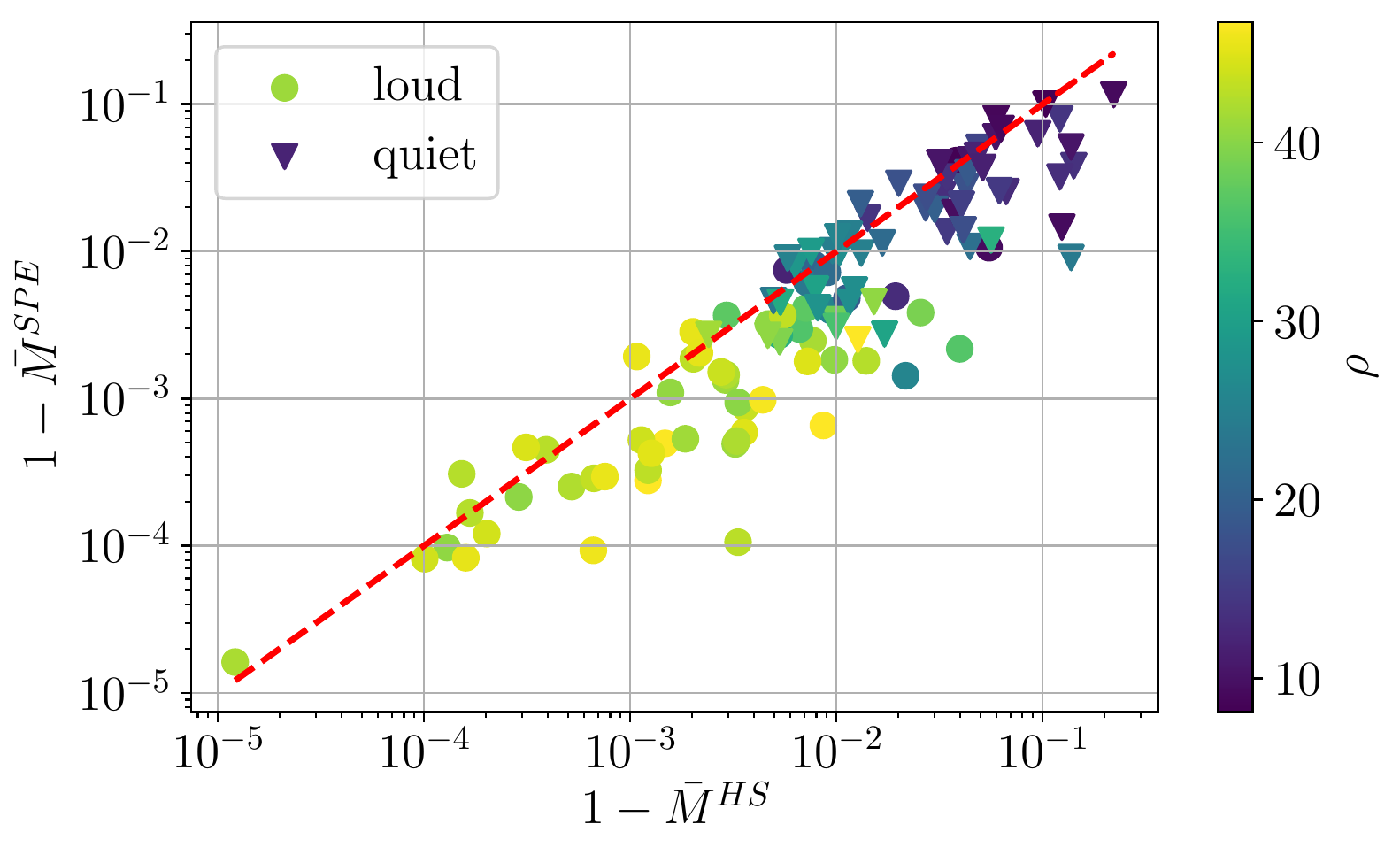}
    \caption{Mismatch for the SPE analysis versus the HS analysis, with noise. The red dashed line represents the diagonal where the mismatch is the same. HS is worse at signal recovery, as most of the points fall below the diagonal.}
    \label{fig:mismatch}
\end{figure}

In Fig.~\ref{fig:mismatch}, we can see mismatches between the injected and recovered waveforms, comparing HS and SPE cases, with and without noise. The mismatch is defined as $1 - \bar{M}$, where $\bar{M}$ is the match between the waveforms, defined as 
\begin{equation}
    \bar{M} = \frac{\langle h_1 | h_2 \rangle}{\sqrt{\langle h_1 | h_1 \rangle\langle h_2 | h_2 \rangle}} \, ,
\end{equation}
where $\langle . | . \rangle$ is the noise weighted product defined in Eq.(\ref{eq:InnerProduct}). The mismatch represents the dissimilarity between two waveforms. High values mean a major disagreement between the two waveforms, and smaller values mean that the waveforms agree well. In Fig.~\ref{fig:mismatch}, we see that the average mismatch throughout the detections is always low, below 0.02. As expected, the presence of another signal leads to worse waveform recovery for HS compared to SPE (most points are below the diagonal). The worst recovery of the signals occurs for overlapping signals with similar SNRs. Note that the zero-noise case (see Appendix~\ref{sec:AppNoNoise}) shows a clearer difference between SPE and HS recoveries. It is expected to have a larger difference in this case because the effect of the unmodelled signal is stronger when there is no noise since it is the only source of uncertainty in the signal.

\begin{figure}
    \centering
    \includegraphics[keepaspectratio, width=0.49\textwidth]{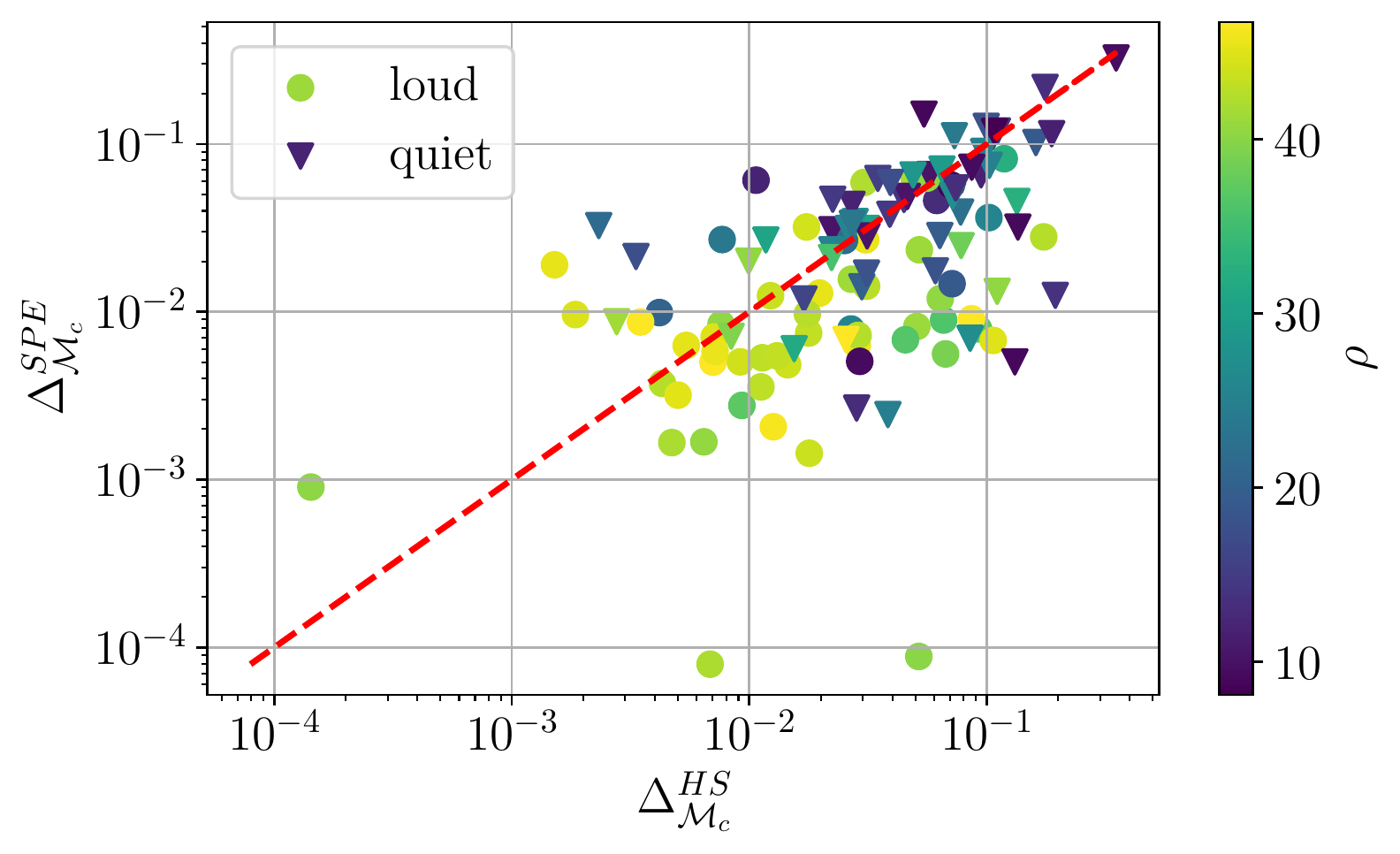}
    \caption{Representation of the offset of the recovered parameters, with noise for HS and SPE. Plotted is the difference between injected value and the median of the recovered value, normalized by the injected value. HS tends to give higher deviations than SPE.}
    \label{fig:biasMcHier}
\end{figure}

By looking at the difference between the median of the recovered posterior and the injected value normalized by the injected value, we can quantify the offset in the recovery ($\Delta_{\mathcal{M}_c} = \frac{|\mathcal{M}_c^{\rm{inj}} - \mathcal{M}_c^{\rm{rec}}|}{\mathcal{M}_c^{\rm{inj}}}$, where ``rec'' stands for recovered and ``inj'' stands for injected.). This is represented in Fig.~\ref{fig:biasMcHier} for the chirp mass recovery with noise and is representative of all the parameters. HS shows a higher offset in 71\% of the louder events and 51\% of the quieter ones. As expected, the deviation is larger for HS compared to SPE. For the first recovered signal, since we have an unmodeled signal present in the detectors, the noise properties are not properly modeled. Therefore, a larger offset is generally observed for the louder signal compared to the secondary one. However, when strong deviations are present for the first signal, it can reverberate on the second, also leading to worse recoveries for this event. 

\begin{figure}
    \centering
    \includegraphics[keepaspectratio, width=0.49\textwidth]{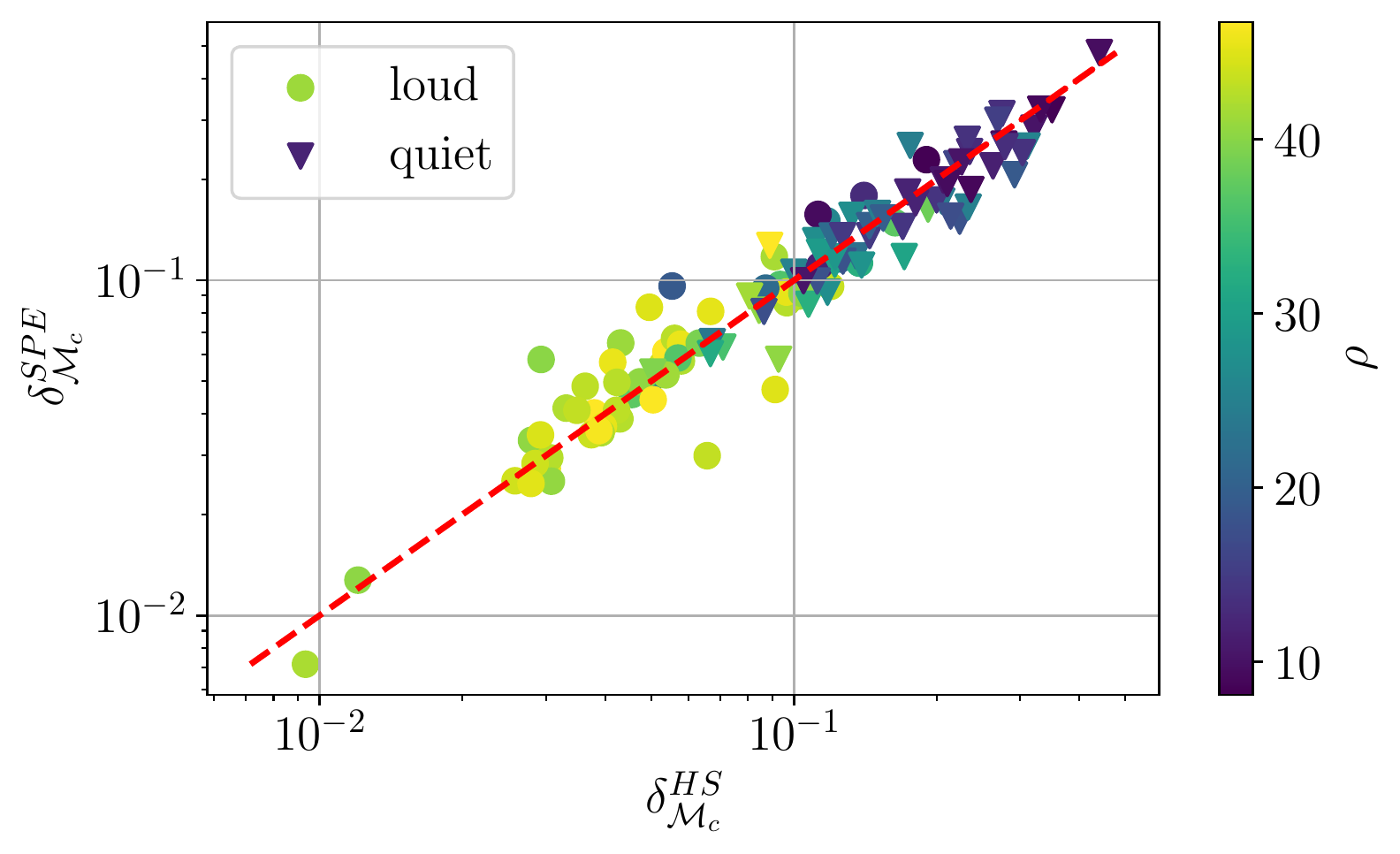}
    \caption{Comparison of the normalized width of the 90\% confidence interval for the chirp mass for the HS recovery and the SPE recovery. The width of the recovered distribution is largely unaffected by the presence of another signal.}
    \label{fig:spreadMcHier}
\end{figure}

Finally, Fig.~\ref{fig:spreadMcHier} shows how the widths of the $90\%$ confidence intervals of the posteriors for HS compare to SPE, normalized by the injected value ($\delta_{\mathcal{M}_c} = \frac{\sigma_{\mathcal{M}_c}}{\mathcal{M}_c^{\rm{inj}}}$, where $\sigma_{\mathcal{M}_c}$ represents the width of the 90\% confidence interval). We observe that the widths of the distribution are consistent between the two, even though the recovery is biased. 

It is of interest to compare how doing successive steps of parameter estimation affects the results. After reconstructing the quieter of the two signals, we subtract it from the initial data and do PE again. In principle, it should result in a better recovery of the louder signal than the original PE run.

\begin{figure}
    \centering
    \includegraphics[keepaspectratio, width=0.49\textwidth]{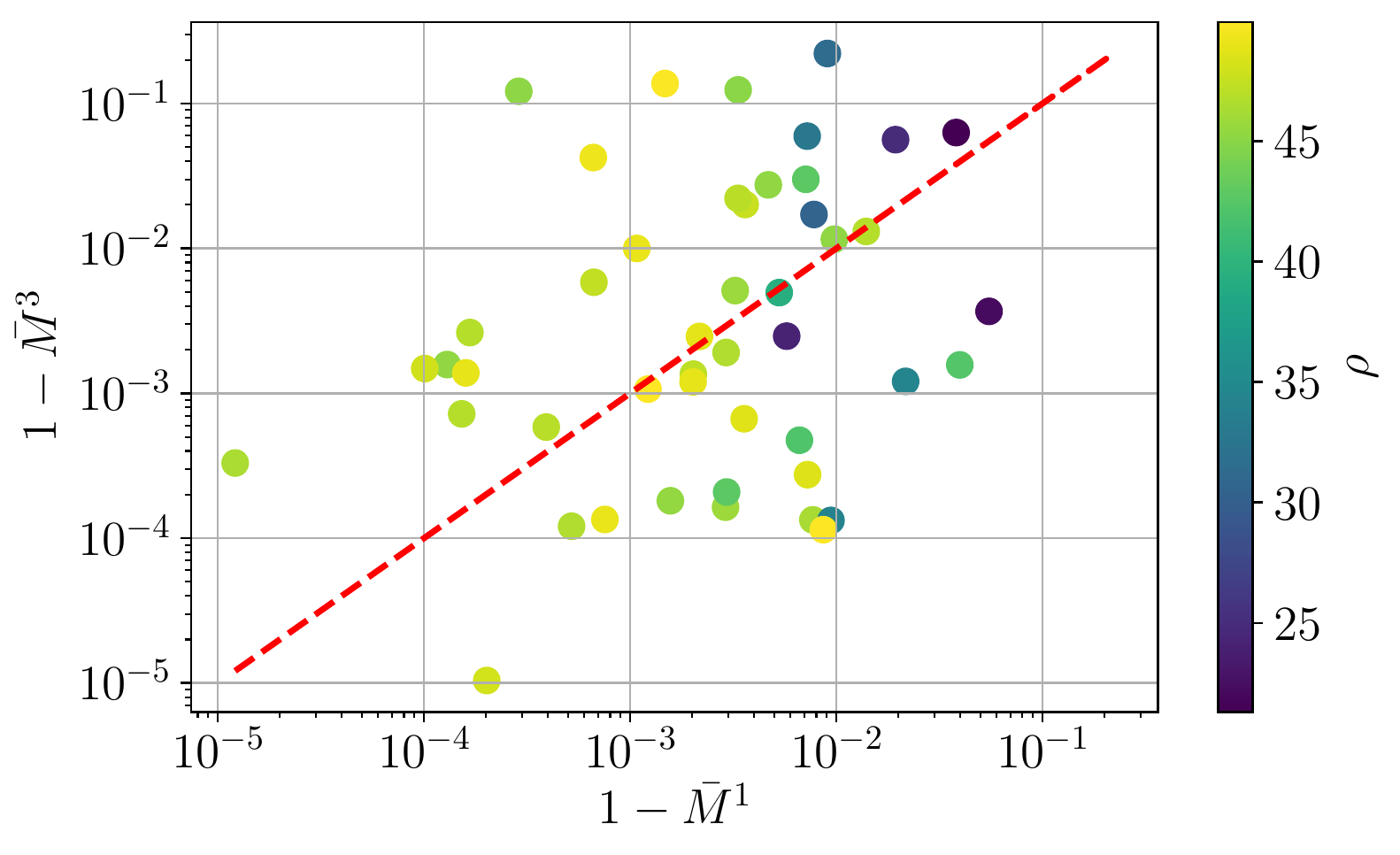}
    \caption{Comparison of the mismatch for the recovered waveform between the first (horizontal axis) and third (vertical axis) steps of HS. The mismatch after the third step is lowered in 62\% of the cases.}
    \label{fig:mismatch1_3}
\end{figure}

\begin{figure}
    \centering
    \includegraphics[keepaspectratio, width=0.49\textwidth]{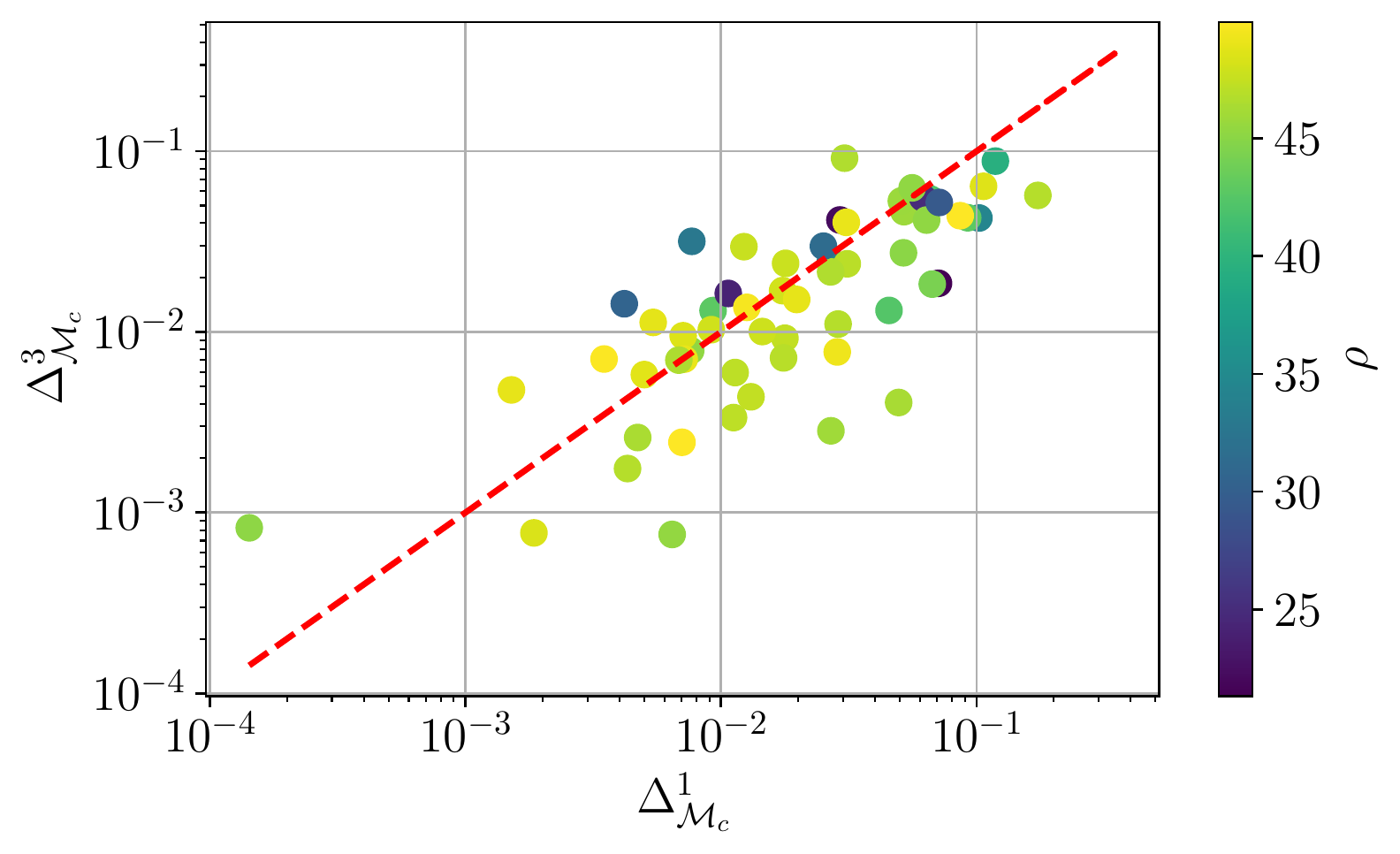}
    \caption{Comparison of the offset in recovered chirp mass between the 1st (horizontal axis) and the 3rd (vertical axis) HS steps. The recovered bias after the 3rd step is lowered in 62\% of the cases.}
    \label{fig:bias1_3}
\end{figure}

Fig.~\ref{fig:mismatch1_3} shows the mismatch for the recovered dominant waveform after the first and third HS run. The match after the third run is better in 62\% of the cases, compared with 50\% expected if the procedure had no effect at all. This small effect is also observed on the offset plot -- see Fig.~\ref{fig:bias1_3} -- where the third HS step leads to better results in the same proportions. Even if the results get better for some events, it also leads to worse recoveries for other cases, and only a few of the other events have comparable results between the first and third HS steps. Therefore, it does not seem like simply applying successive HS steps converges to posteriors unaffected by the overlap. More sophisticated approaches appear to be needed, like an estimation of the deviation as suggested in~\citet{Antonelli:2021vwg} and briefly explained in Section~\ref{sec:method_hier}.

In the end, our runs for HS confirm what has been seen in previous research~\citep{Samajdar:2021egv, Relton:2021cax, Pizzati:2021apa}: doing PE for one signal neglecting the other can lead to significant biases when the two signals merge very close to each other. In addition, we have also shown that once the first signal is subtracted, analyzing the second one with parameter estimation is less prone to deviations, even if the subtraction of the first event is not perfect.

\subsection{Joint parameter estimation}

In what follows, we will discuss the results obtained from the analysis of the overlapping signals using JPE. We focus on the results obtained after having done the time ordering of the samples described in Sec.~\ref{sec:methods}, as these are the samples where effectively one set of posteriors is matched with one signal, and the other set is matched with the other signal. 

\begin{figure}
    \centering
    \includegraphics[keepaspectratio, width=0.4\textwidth]{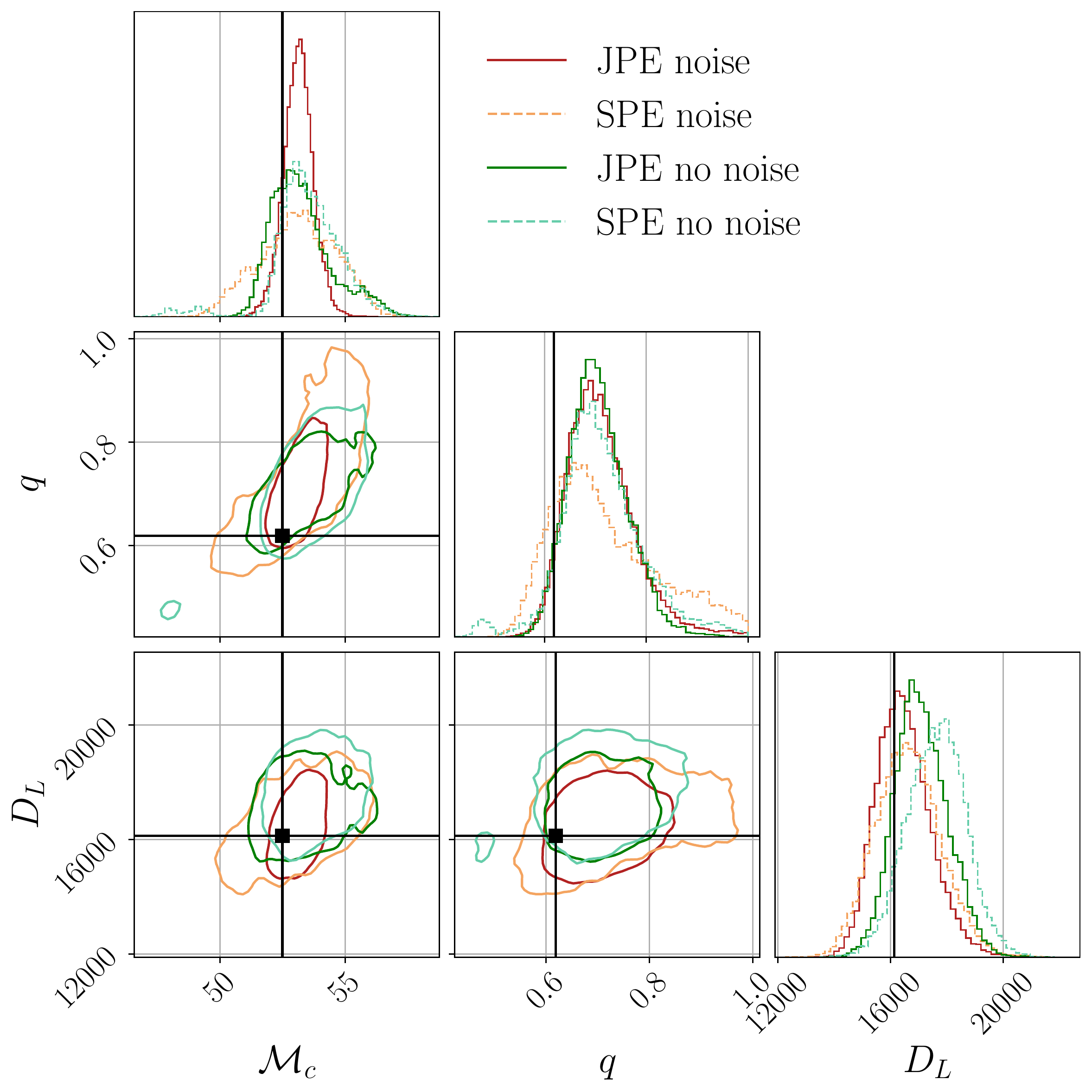}
    \includegraphics[keepaspectratio, width=0.4\textwidth]{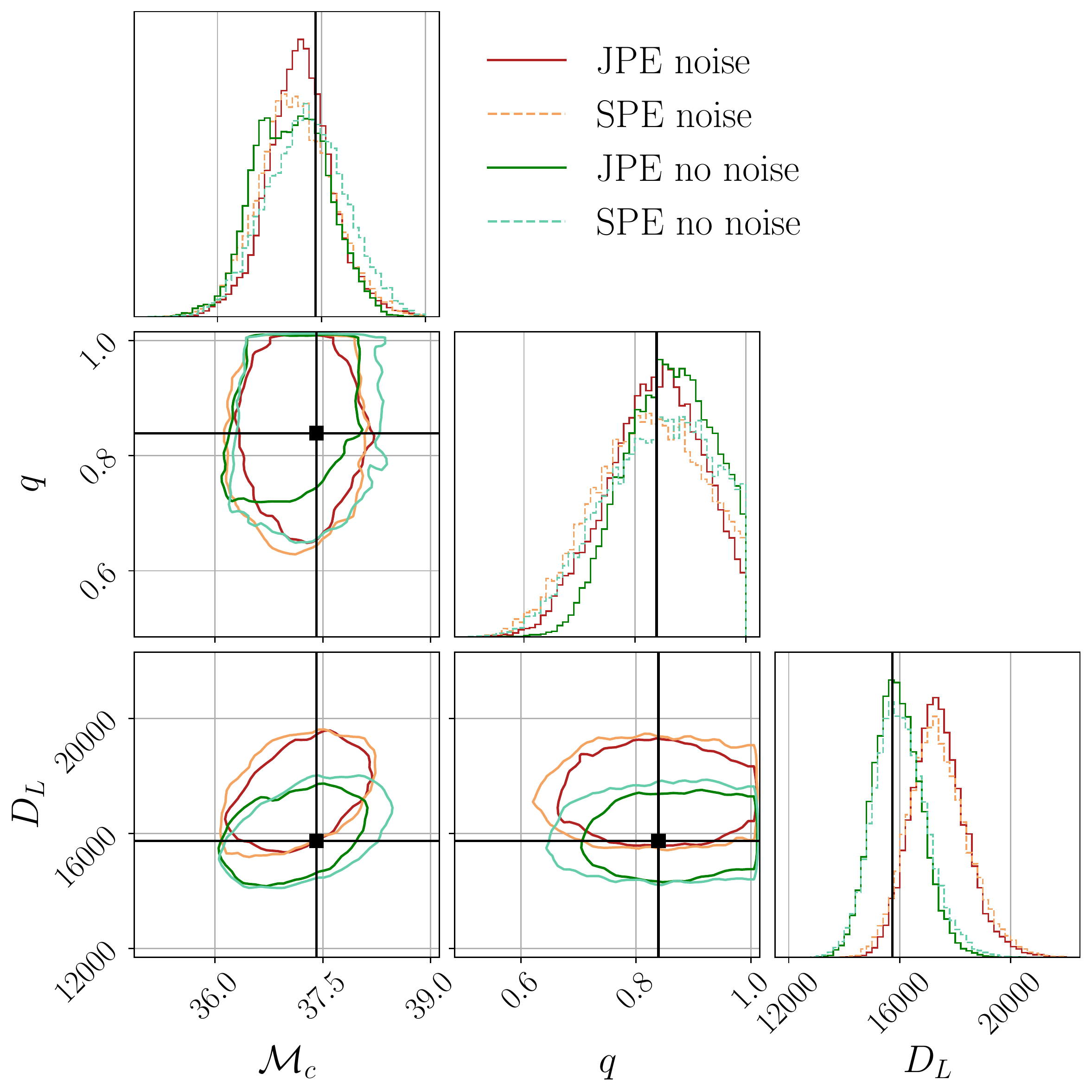}
    \includegraphics[keepaspectratio, width=0.4\textwidth]{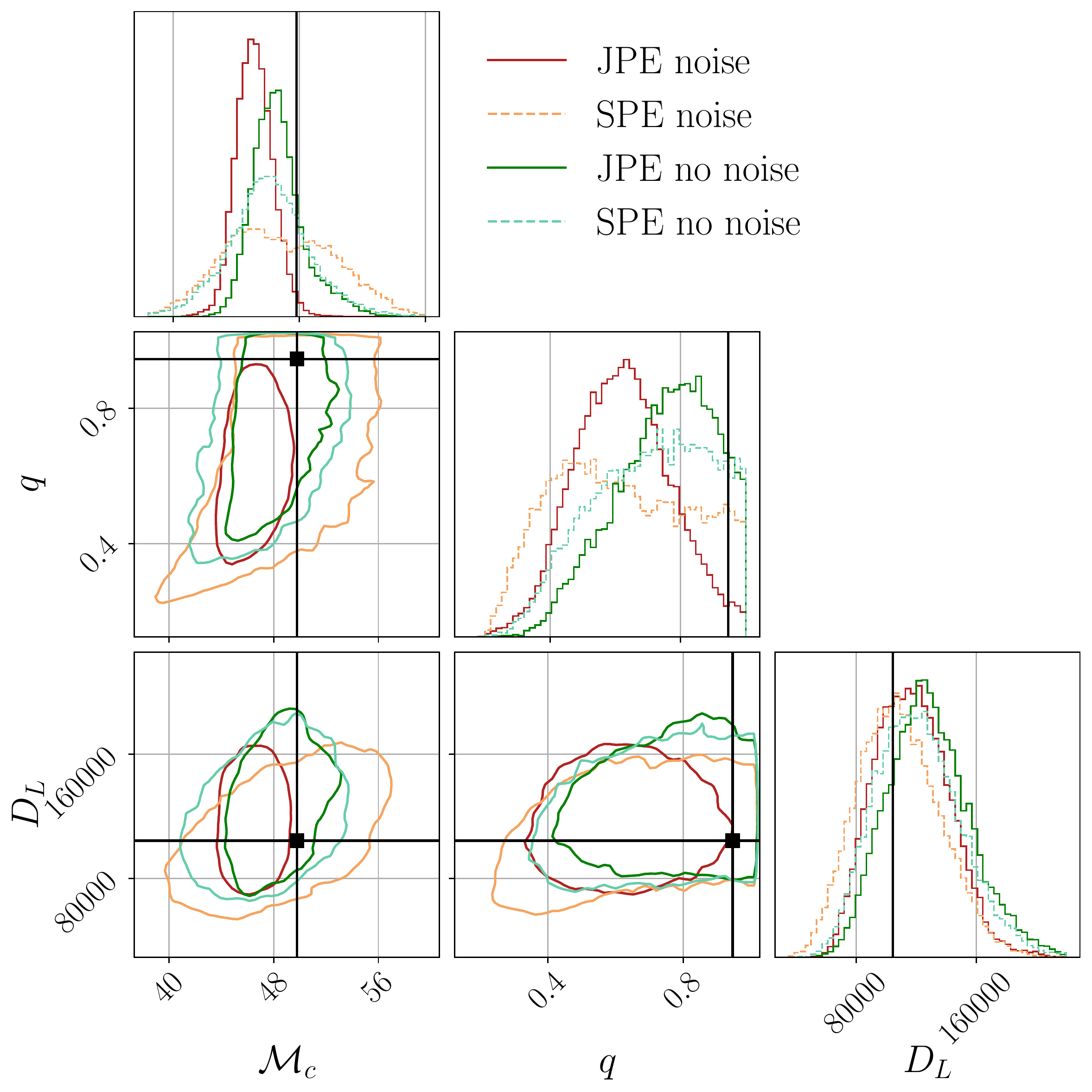}
    \caption{Comparison of the posteriors recovered for JPE and SPE with and without noise for different types of recovery. \emph{Top:} the posteriors recovered with JPE for the noise case are narrower for the chirp mass and the mass ratio compared to the single PE case, while the posterior for the luminosity distance is narrower for JPE in the zero-noise case. \emph{Middle:} JPE and SPE are very close to each other, with equivalent recovery in the two cases. \emph{Bottom:} representation of a case where the JPE recovery is worse than for SPE. We get narrower posteriors, but the peak is shifted out of the 90\% confidence region.}
    \label{fig:JPEposteriors}
\end{figure}

For the JPE recovery, more diverse scenarios are possible. We show 3 main cases in Fig.~\ref{fig:JPEposteriors}: one where the recovery is equivalent to the one from SPE, one where JPE has smaller bounds on the posteriors for one or more parameters, and one where JPE has trouble fitting the signal properly and biases can occur. While more in-depth studies are required to fully comprehend this behavior, it could be originating from the mixed term of the two signals present in the likelihood when it was modified to account for multiple signals. In some cases, the narrower posteriors could be offset compared to the injected value, pushing the latter out of the 90\% confidence interval.

\begin{figure*}
    \centering
    \includegraphics[keepaspectratio, width=0.49\textwidth]{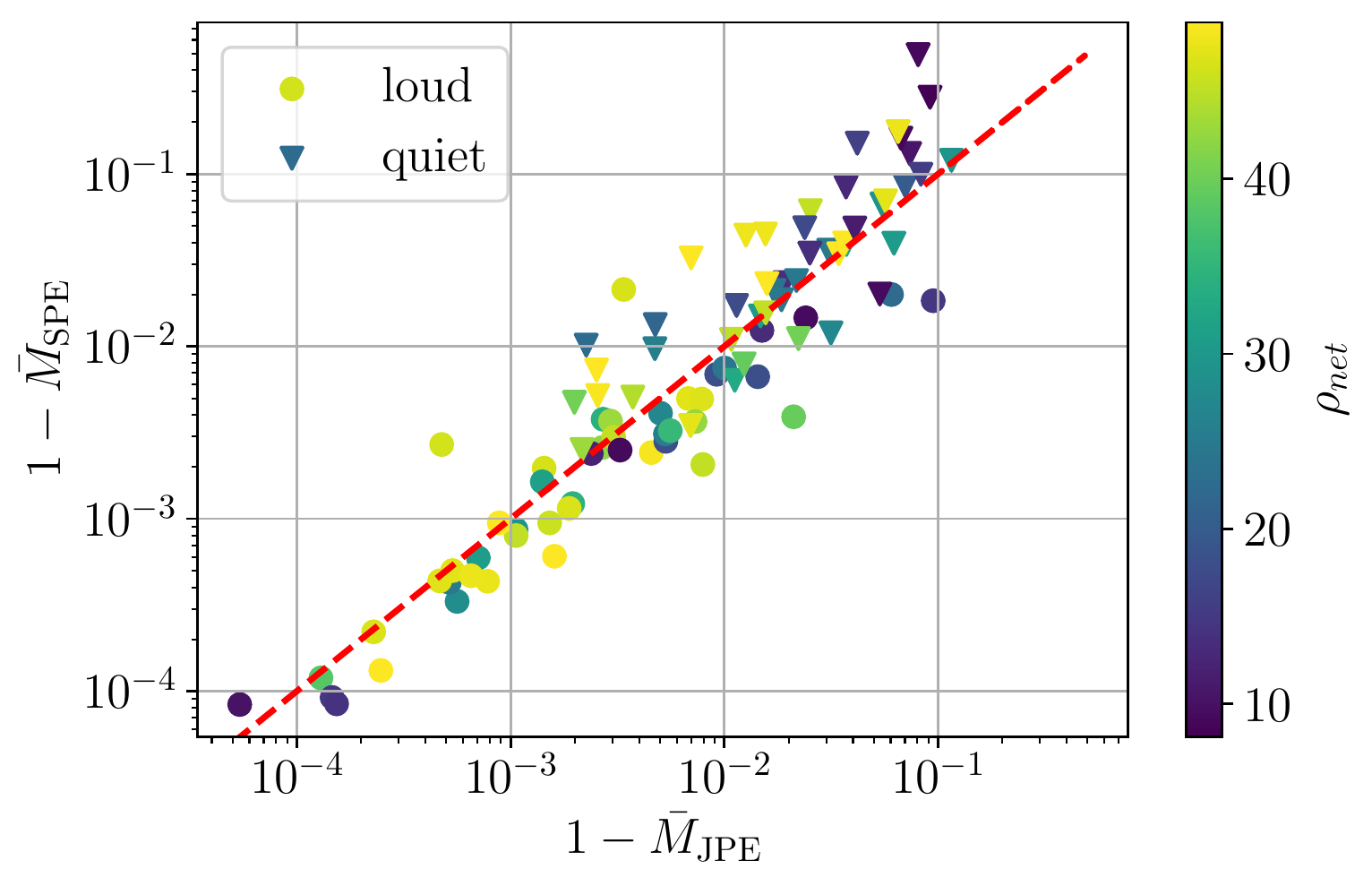}
    \includegraphics[keepaspectratio, width=0.49\textwidth]{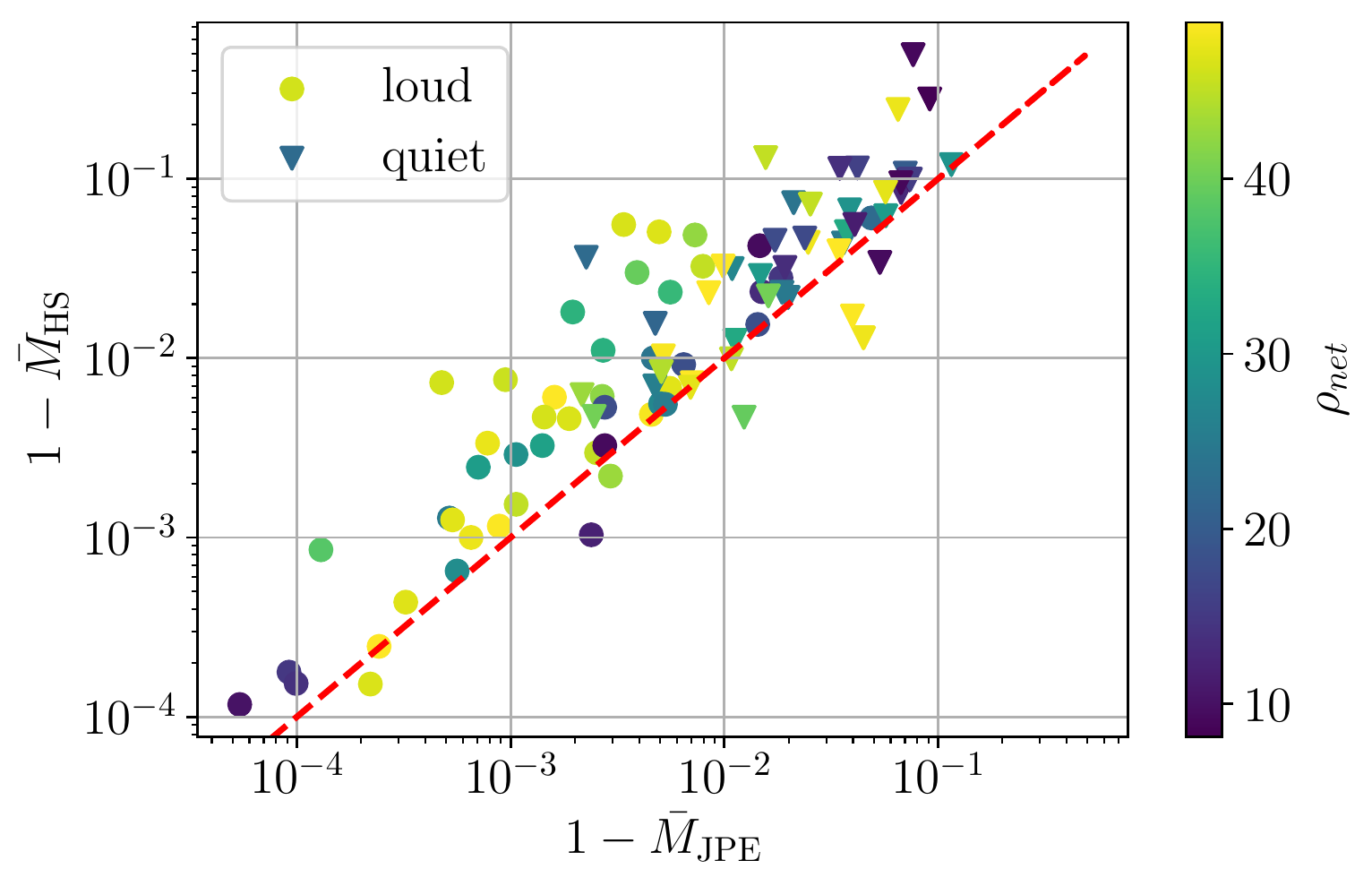}
    \caption{\emph{Left:} comparison of the mismatch for the JPE and the SPE cases with noise. \emph{Right:} comparison of the mismatch for JPE and HS with noise. We see that overall, the mismatch is higher for JPE than for SPE, while it is lower than for the HS case. This is expected since JPE accounts for the two events in the data, which is better than neglecting one but more complicated than having only one signal present in the data and fitting that signal.}
    \label{fig:JPEmatches}
\end{figure*}

For all the events, we compare the mismatch between the maximum likelihood waveform for the event in the JPE scenario with the maximum likelihood for the event in the SPE and HS cases. This is represented in Fig.~\ref{fig:JPEmatches} for the noise cases. Independent of the presence of noise, we find that the mismatch is smaller for JPE than for HS but larger than for SPE. This is what one would expect since JPE accounts for the presence of the two signals and so should lead to smaller biases. However, fitting simultaneously two signals is more complex than analyzing a single signal. Therefore, the SPE measures remain a better representation of the injected signal. 

\begin{figure*}
    \centering
    \includegraphics[keepaspectratio, width=0.49\textwidth]{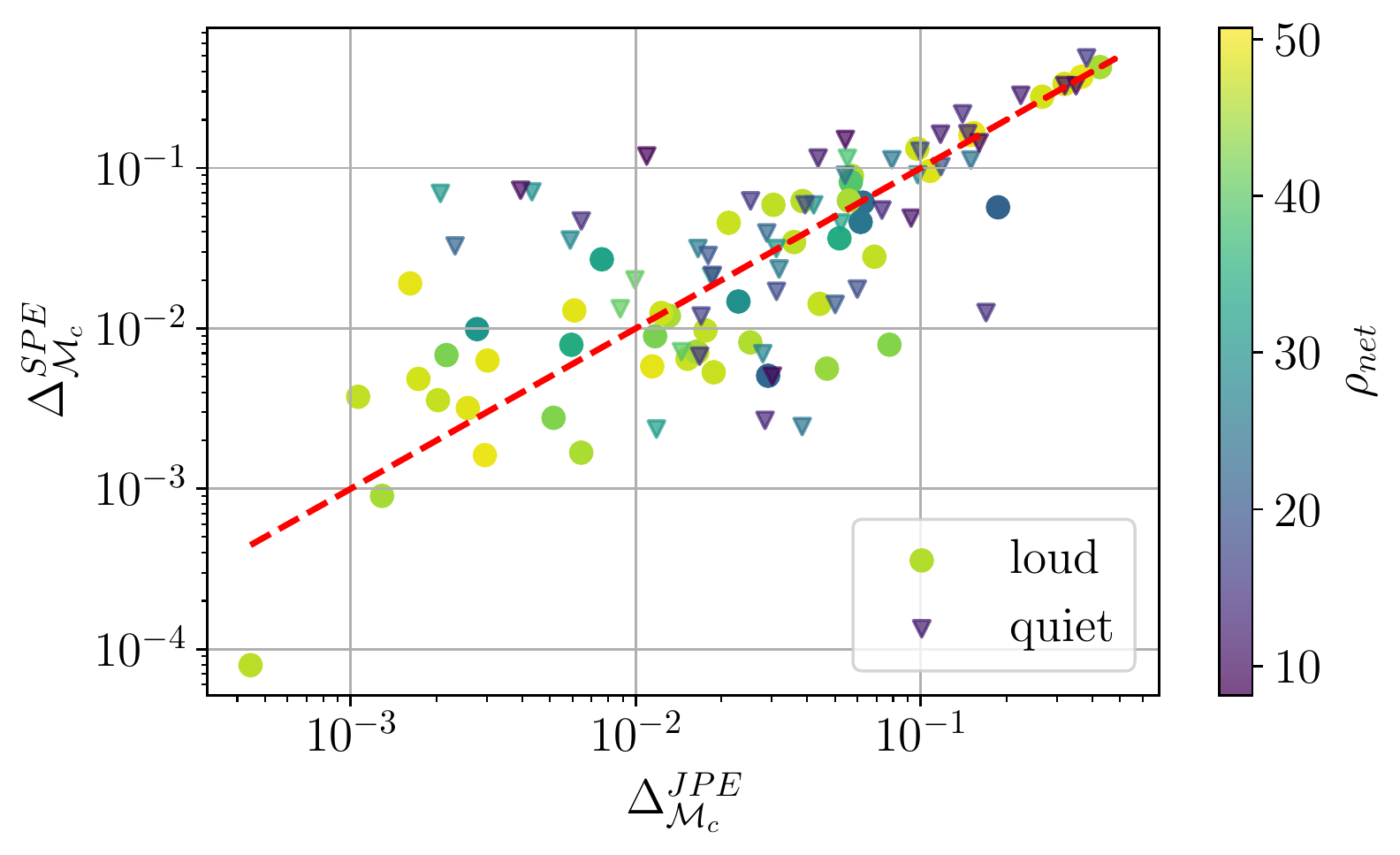}
    \includegraphics[keepaspectratio, width=0.49\textwidth]{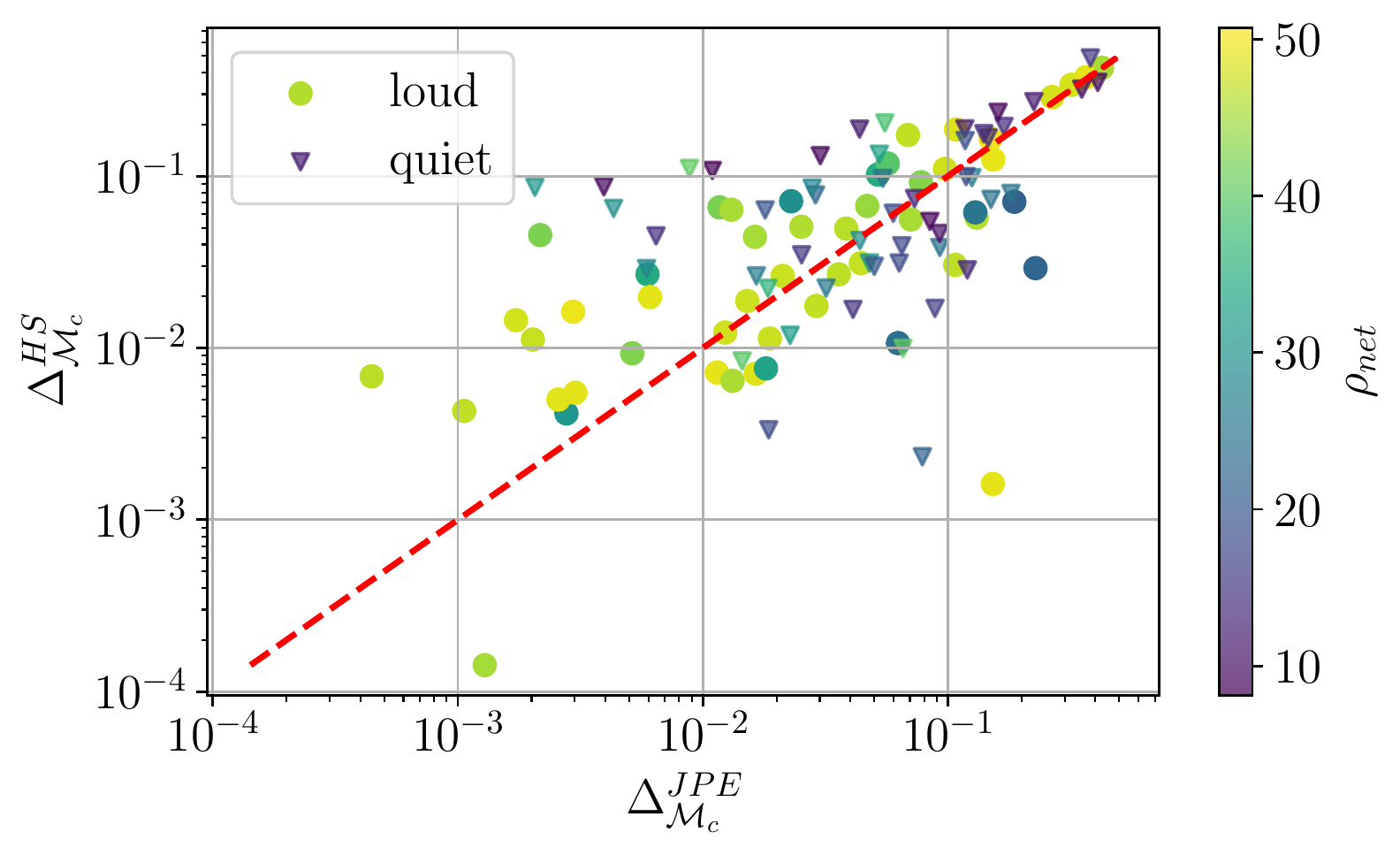}
    \caption{\emph{Left:} comparison of the offset of the recovered posterior for the chirp mass for JPE and SPE method. \emph{Right:} comparison of the offset of the recovered posteriors for the chirp mass for JPE and HS. The two plots indicate that the offset is reduced for JPE compared to HS, due to the better modeling of the noise, while it is still better in the SPE case, where the noise is well modeled and the problem at hand has a reduced complexity.}
    \label{fig:biasJPEnoise}
\end{figure*}

As before, we present the normalized distance between injection and recovery in Fig.~\ref{fig:biasJPEnoise} for the noise case. When comparing this offset for the JPE case against the SPE case, we find that 45\% of the events have a lower offset for JPE than for SPE. On the other hand, when we compare with HS, we find 65\% of the events with a lower offset for JPE. This confirms that JPE is better than HS to find the injected signals (when two signals are present at $0.1\rm s$ of each other in the data). This is indeed what one would expect, as JPE takes care of the mismodeling of the noise but leads to an increased complexity during the analysis. 

\begin{figure}
    \centering
    \includegraphics[keepaspectratio, width=0.49\textwidth]{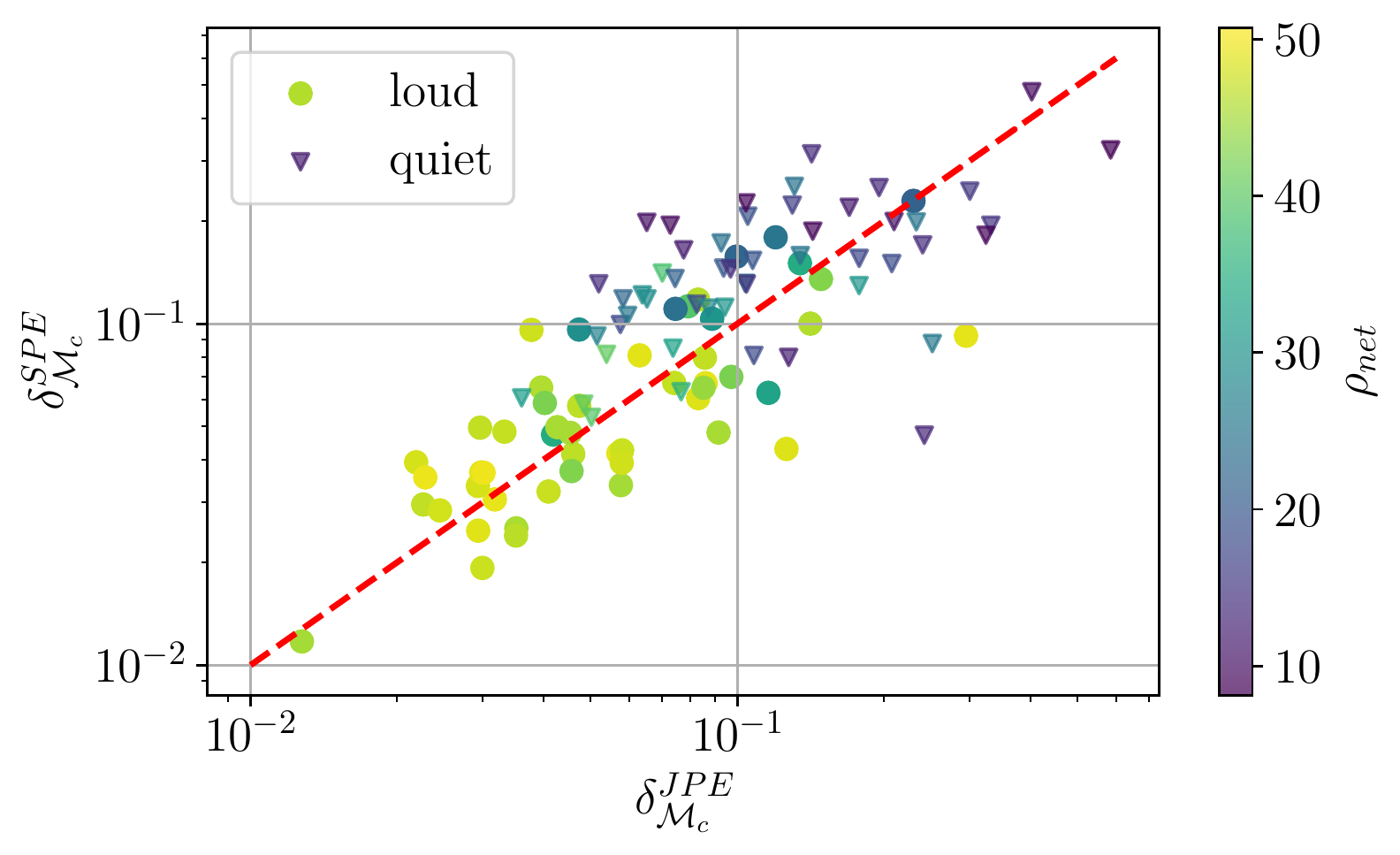}
    \caption{Comparison between the normalized width of the 90\% confidence interval for JPE and SPE for the noise case. Since the spreads are very close for SPE and HS, the same relation is valid for the JPE and HS comparison. We see that there are larger differences here between the two approaches but that globally, one is not better than the other as we have about 50\% of the events above the diagonal and the same proportion below.}
    \label{fig:SpreadJPEnoise}
\end{figure}

For the spread in recovered posteriors, contrary to what one had for the HS approach, the normalized width of the 90\% confidence interval does not align itself on the diagonal. Indeed, since we have more varying scenarios, with larger or tighter posteriors in some cases, the spread can be significantly larger or tighter in the JPE case compared to the SPE scenario. In addition, since there is no significant difference for this quantity between HS and SPE, the relation between JPE and HS is the same as between JPE and SPE. The increased discrepancy between the two approaches is represented in Fig.~\ref{fig:SpreadJPEnoise}. Nevertheless, the posteriors are evenly distributed above and below the diagonal representation, showing that, on average, the width of the posteriors is not significantly bigger in one method or the other. 

Based on our JPE results, we see that, using an adapted likelihood, we obtain better results than by analyzing both signals sequentially. However, this comes with the drawback that the computational time and the complexity of the problem are increased, making the approach less stable as can be seen by the wider variations in the widths of the posteriors. On the other hand, in some cases, JPE leads to narrower posteriors compared to SPE. This could be due to the inner product from the mixed term of the two signals following the introduction of multiple signals in the likelihood. However, a more in-depth study should be performed on a larger set of events to understand this behavior. This is left for future work.

\section{Conclusions}
\label{sec:conclusions}

In this work, we have presented two ways to perform parameter estimation for two overlapping BBH signals: hierarchical subtraction and joint parameter estimation. We have implemented them and compared them to the usual single-parameter estimation. Because of limited computational resources, we focus on high mass and medium SNR events; major adaptions to the parameter estimation framework are needed to deal with a larger variety of signals.

For the hierarchical subtraction method, we analyze the first signal, then subtract it (more precisely, the maximum likelihood signal) from the data before analyzing the second signal still left in the data. On the other hand, joint parameter estimation consists of analyzing the two signals simultaneously. We note that since the likelihood is symmetric in the two signals when they have the same nature, we need to add a post-processing step to have samples corresponding to each event. For this purpose, we order the samples in time.

We have applied both methods to a population of binary black hole mergers to show the feasibility of the two approaches and their respective drawbacks. For hierarchical subtraction,  as already mentioned in the literature, we showed that analyzing one signal while neglecting the other can lead to offsets in the recovered posteriors because of the erroneous noise representation. However, even when some deviations occur in the first signal, it does not necessarily impact the recovery of the second signal. We have also shown that there is no significant broadening of the posteriors compared to usual single-parameter estimation approaches. Therefore, it shows that hierarchical subtraction suffers from biases due to overlapping signals for signals measured at very close merger times, which can lead to an incorrect inference of the parameters. 

For joint parameter estimation, we have shown that the offset in recovered posteriors is smaller than for hierarchical subtraction while remaining higher than for single parameter estimation. This is understood as joint parameter estimation having a correct noise representation for the two signals, which hierarchical subtraction lacks. On the other hand, solving the likelihood for joint parameter estimation means we need to explore a 30-dimensional parameter space, making the analysis more complex and computationally challenging than single parameter estimation. However, the recovered width of the posteriors is, on average, approximately the same as for single parameter estimation. But while the average is the same, we are confronted with posteriors that can be narrower or broader in joint parameter estimation compared to single parameter estimation. This behavior may originate from the cross-term between the two signals entering the likelihood when adapting it to account for multiple signals. More extended studies are needed to understand this effect.

Overall, our results indicate that common techniques for a joint likelihood approach are not yet at their best, and several options to make the sampling more efficient could be possible for future work. For example, one could impose the time ordering (or chirp mass hierarchy) directly during the sampling, by imposing that the arrival time of one event is smaller than that of another event. This would prevent the sampler from confusing the two events and enable them to converge more easily. Another possibility could be to use narrower priors motivated by the output of the search pipelines. This is possible for the chirp mass and the time of arrival but still contains risk as the search pipelines themselves can provide inaccurate point estimates for some critical parameters~\citep{Relton:2022whr}.

One of the major issues with the methods suggested here is the computational time required, as the data analysis takes up to a few months for overlapping binary black hole mergers with lower component masses. This would make it extremely hard, if not impossible, to keep up with the detection rate of the 3G-detector network. However, methods exist to speed up traditional parameter estimation methods, such as relative binning~\citep{Dai:2018dca, Zackay:2018qdy, Leslie:2021ssu} or adaptive frequency resolution~\citep{Morisaki:2021ngj}. These methods could be adapted to overlapping signals in future work to reduce the computational time, enabling one to analyze other types of systems and to use a lower minimum frequency to get closer to the real 3G scenario. A totally different approach that could help in the analysis of such signals in the future is machine learning, where major progress has been made in the parameter inference for single compact binary colescences~\citep{Dax:2021tsq, Kolmus:2021buf}. In the future, one could think of adapting these methods to overlapping signals. In parallel to this work, a proof-of-concept study applying machine-learning techniques to overlapping signals has shown that it is possible to extract posteriors using normalizing flows with a reasonable precision~\citep{Langendorff:2022fzq}.

To this end, we believe that this work makes a first step towards the analysis of overlapping compact binary coalescence signals, which will be crucial to analyze gravitational wave data in the third-generation detector era, as overlaps will become quite common. \\

\section*{Acknowledgments}
The authors thank Philip Relton for useful comments on the manuscript.
J.J., T.B., and C.V.D.B are supported by the research program of the Netherlands Organisation for Scientific Research (NWO). 
A.S. thanks the Alexander von Humboldt foundation in Germany for a Humboldt fellowship for postdoctoral researchers. 
The authors are grateful for computational resources provided by the LIGO Laboratory and supported by the National Science Foundation Grants No. PHY-0757058 and No. PHY-0823459. We are grateful for computational resources provided by Cardiff University, and funded by an STFC grant supporting UK Involvement in the Operation of Advanced LIGO

\section*{Data availability}
The data underlying this article will be shared in reasonable request to the corresponding authors.

\bibliography{bibli}

\newpage
\appendix
\section{Zero-noise results}
\label{sec:AppNoNoise}
In this section, we show results for the posteriors obtained when the injections are analyzed without noise. The conclusions drawn from these plots are the same as in the noise case, which suggests that our findings are robust, due to sampling effects, and not induced by the random noise added to the data.

\subsection{Hierarchical Subtraction}
In this section, we present the complementary zero-noise result for the HS method.

\begin{figure}
    \centering
    \includegraphics[keepaspectratio, width=0.49\textwidth]{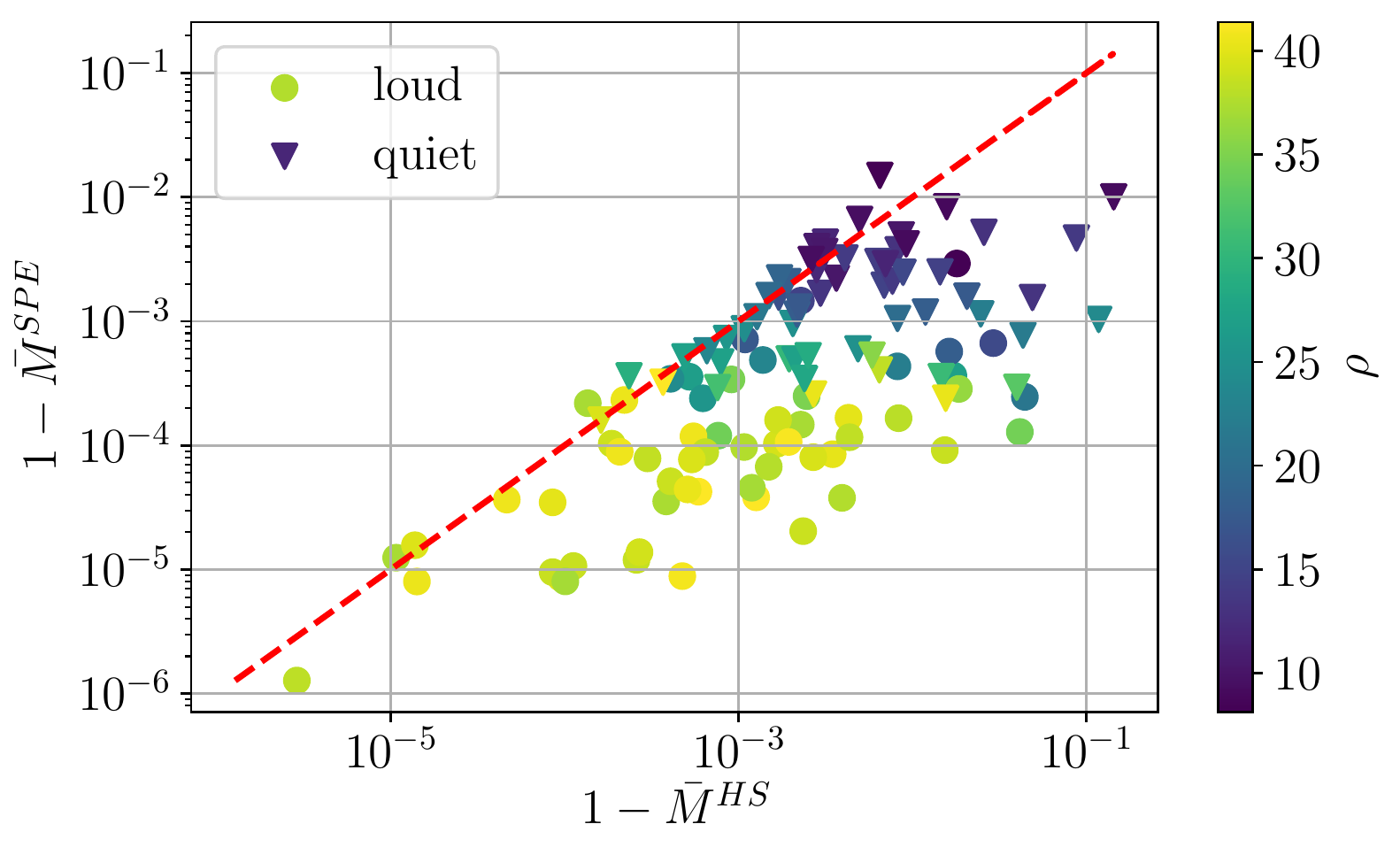}
    \caption{Mismatch for the SPE analysis versus HS analysis without noise. The red dashed line represents the diagonal where the mismatch is the same. The difference in waveform recovery between SPE and HS is more pronounced than in the case with noise, and it is clear that the recovery is worsened when using the HS approach.}
    \label{fig:mismatch_no_noise}
\end{figure}

Fig.~\ref{fig:mismatch_no_noise} represent the mismatch for HS versus SPE for the zero-noise case. As for the noise case,  HS leads to higher mismatches, meaning that the recovered (maximum likelihood) parameters are a worse representation of the injected signals.

\begin{figure}
    \centering
    \includegraphics[keepaspectratio, width=0.49\textwidth]{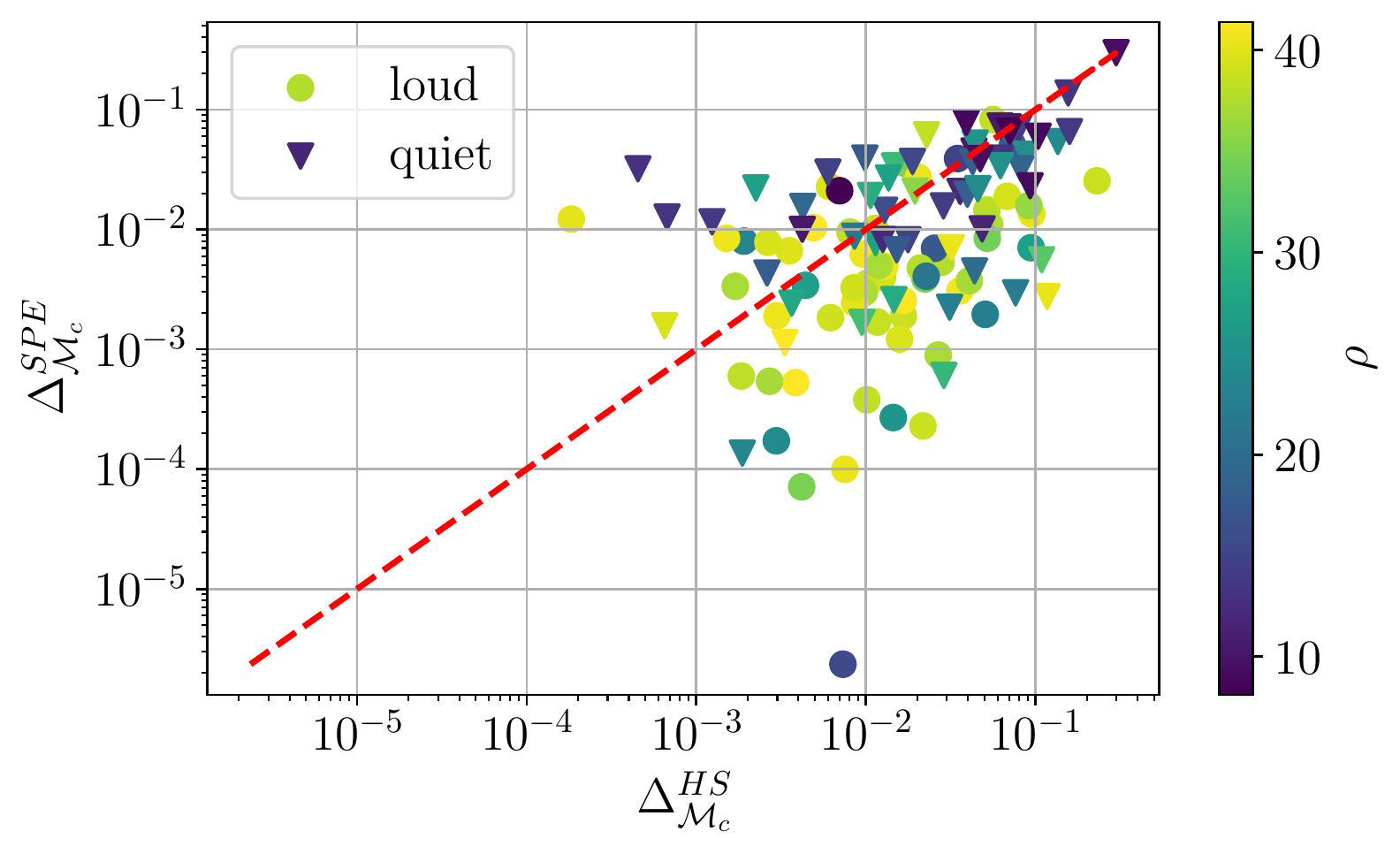}
    \caption{Representation of the offset of the recovered chirp mass without noise. Plotted is the difference between injected value and the median of the recovered value, normalized by the injected value for HS and SPE. One sees that offsets are more important for HS.}
    \label{fig:biasMcHier_no_noise}
\end{figure}

Fig.~\ref{fig:biasMcHier_no_noise} represent the offset for the chirp mass for HS versus SPE without noise. Here HS shows higher offset for 74\% of the louder events and 57\% of the quieter ones. There is no significant difference compared with the injections into noise, and HS is still prone to more deviations. 

\begin{figure}
    \centering
    \includegraphics[keepaspectratio, width=0.49\textwidth]{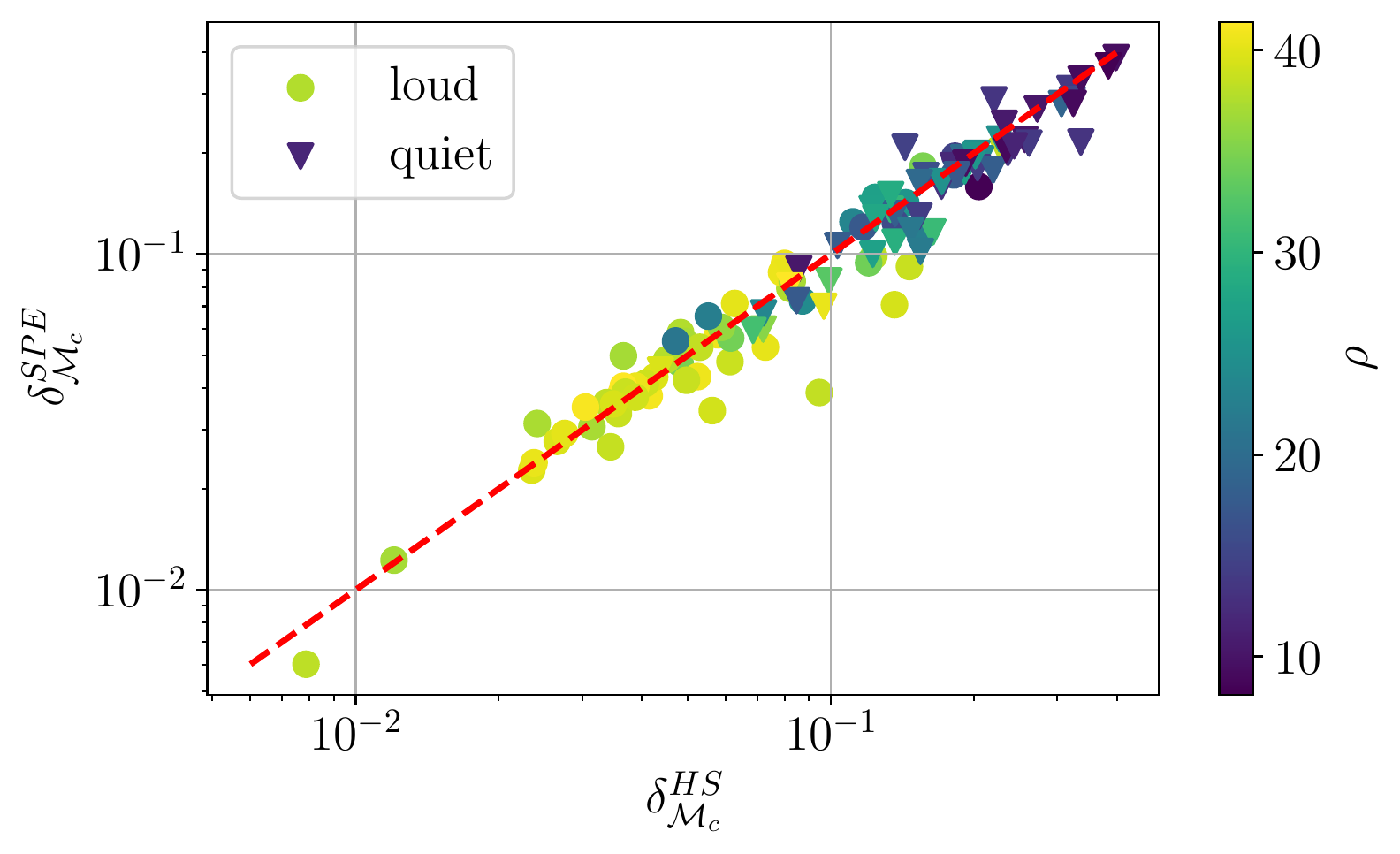}
    \caption{Comparison of the normalized width of the 90\% confidence interval for the chirp mass for the HS recovery and the SPE recovery, with zero-noise. The width of the recovered distribution is largely unaffected by the presence of another signal.}
    \label{fig:spreadMcHier_no_noise}
\end{figure}

Fig.~\ref{fig:spreadMcHier_no_noise} represents the normalized width of the posteriors for HS versus SPE. Similarly to the analysis with noise, the widths of the distributions are very close to each other.

\subsection{Joint Parameter Estimation}
In this section, we present the complementary zero-noise results for the JPE method. 

\begin{figure*}
    \centering
    \includegraphics[keepaspectratio, width=0.48\textwidth]{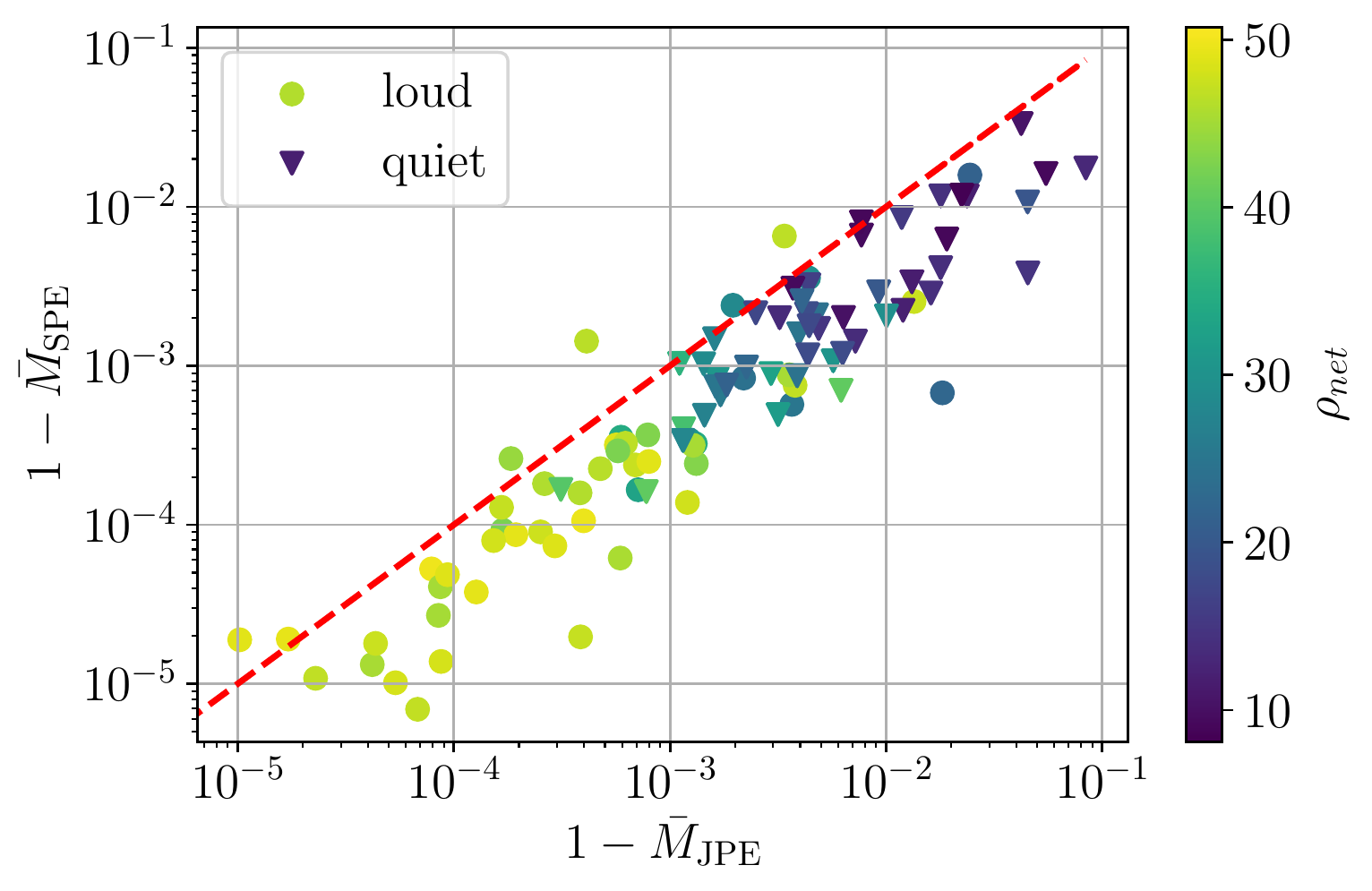}
    \includegraphics[keepaspectratio, width=0.48\textwidth]{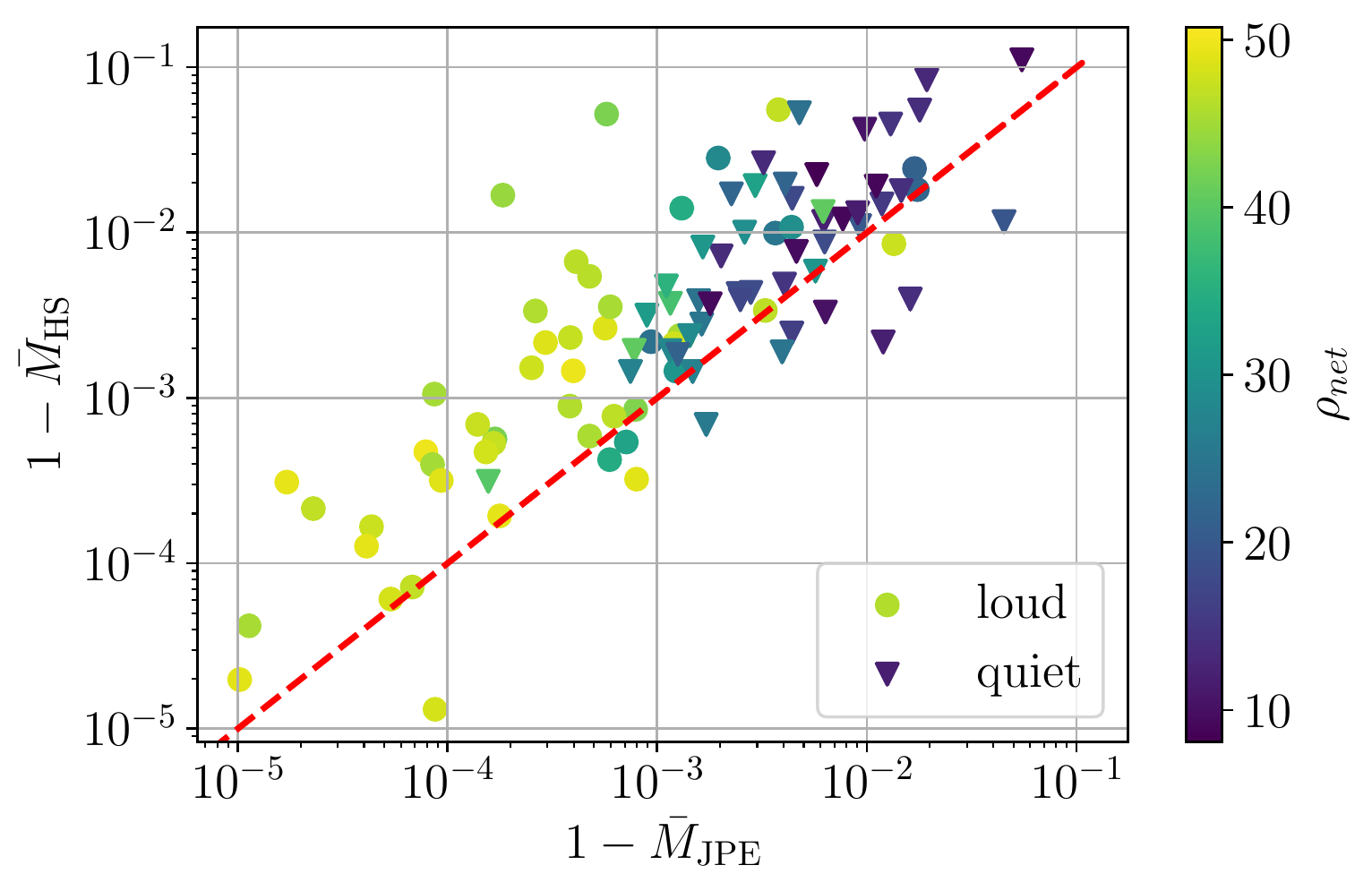}
    \caption{\emph{Left:} comparison of the mismatch for the recovered events for the two events for JPE and SPE without noise. \emph{Right:} comparison of the mismatch for JPE and HS without noise. Overall, the mismatch is higher for JPE than for SPE, while it is lower than for the HS approach. This is expected since JPE accounts for the two events in the data, which is better than neglecting one but more complicated than having only one signal present in the data and fitting that signal.}
    \label{fig:JPEmatchesNoNoise}
\end{figure*}

Fig.~\ref{fig:JPEmatchesNoNoise} represent the mismatch between the JPE and SPE methods (left) and the JPE and the HS methods (right). This also shows that the JPE method leads to a better representation of the data than the HS method, but that the increased complexity of the problem leads to a decrease in the accuracy of the recovery. Still, the mismatch values are relatively low in the two cases. 

\begin{figure*}
    \centering
    \includegraphics[keepaspectratio, width=0.48\textwidth]{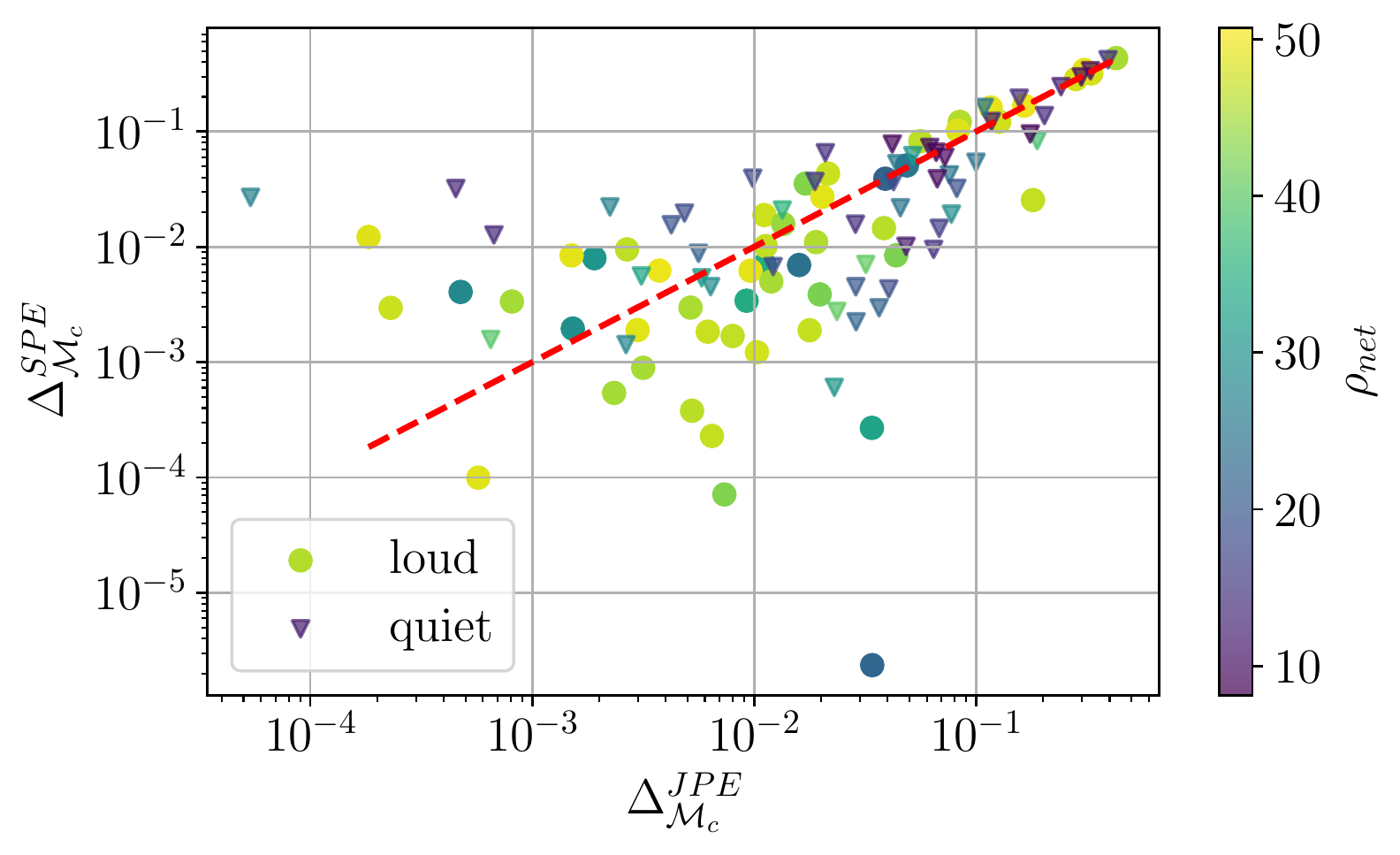}
    \includegraphics[keepaspectratio, width=0.48\textwidth]{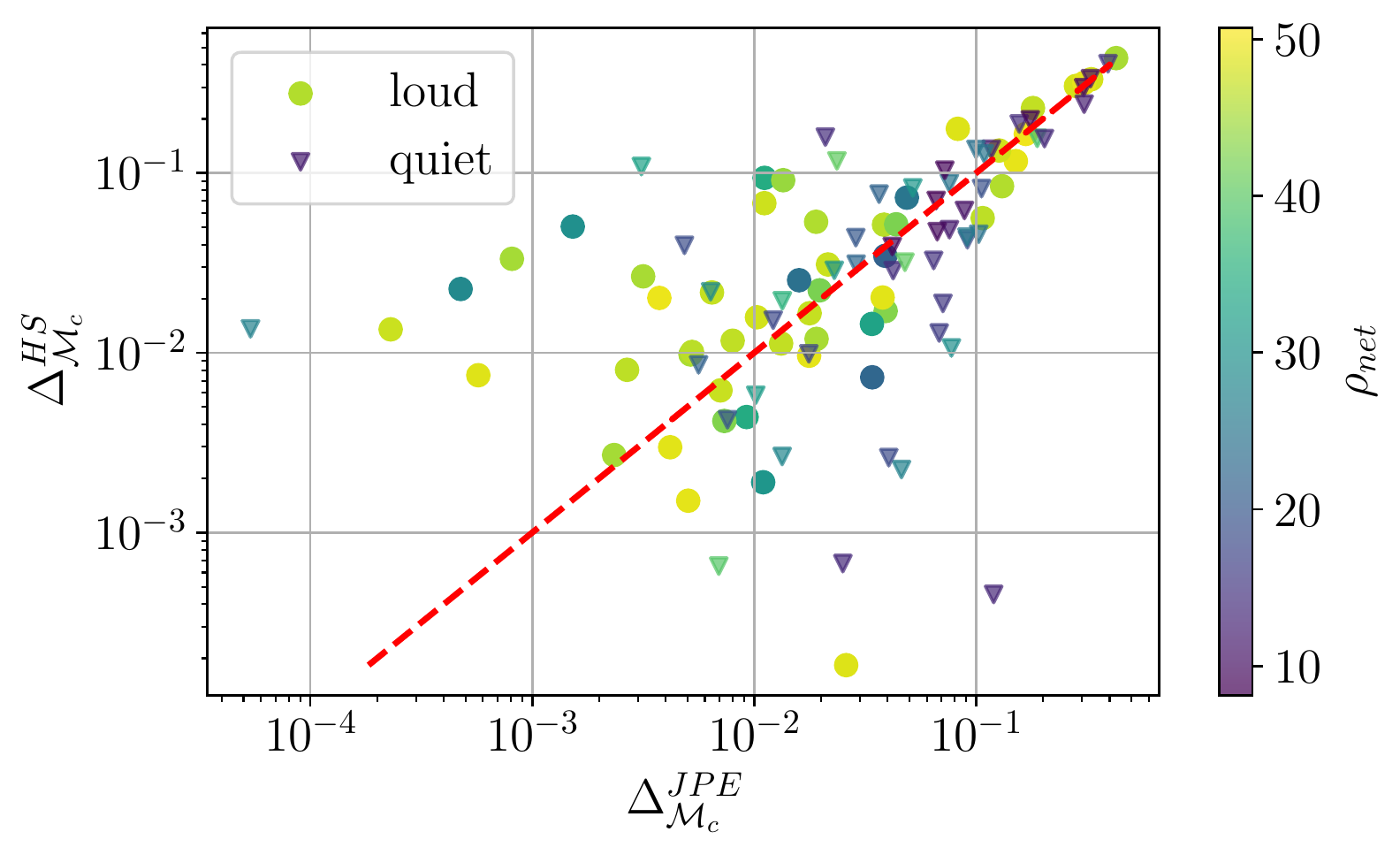}
    \caption{\emph{Left:} comparison of the offsets of the recovered posterior for the chirp mass for JPE and SPE method in the zero-noise case. \emph{Right:} comparison of the offsets of the recovered posteriors for the chirp mass for JPE and HS in the zero-noise case. The two plots indicate that the offset is lower for JPE than for HS, due to the better modeling of the noise, while it is still better in the SPE case, where the noise is well modeled and the problem at hand has a reduced complexity.}
    \label{fig:biasJPEnoNoise}
\end{figure*}

Fig.~\ref{fig:biasJPEnoNoise} represent the comparison between the offsets between JPE and SPE, and JPE and HS for the zero-noise case. Here, one also has a larger offset for the JPE than for SPE (39\% of the events have smaller offset for JPE), and a larger offset for HS as for JPE (57\% of the events have a smaller offset for JPE).

\begin{figure}
    \centering
    \includegraphics[keepaspectratio, width=0.49\textwidth]{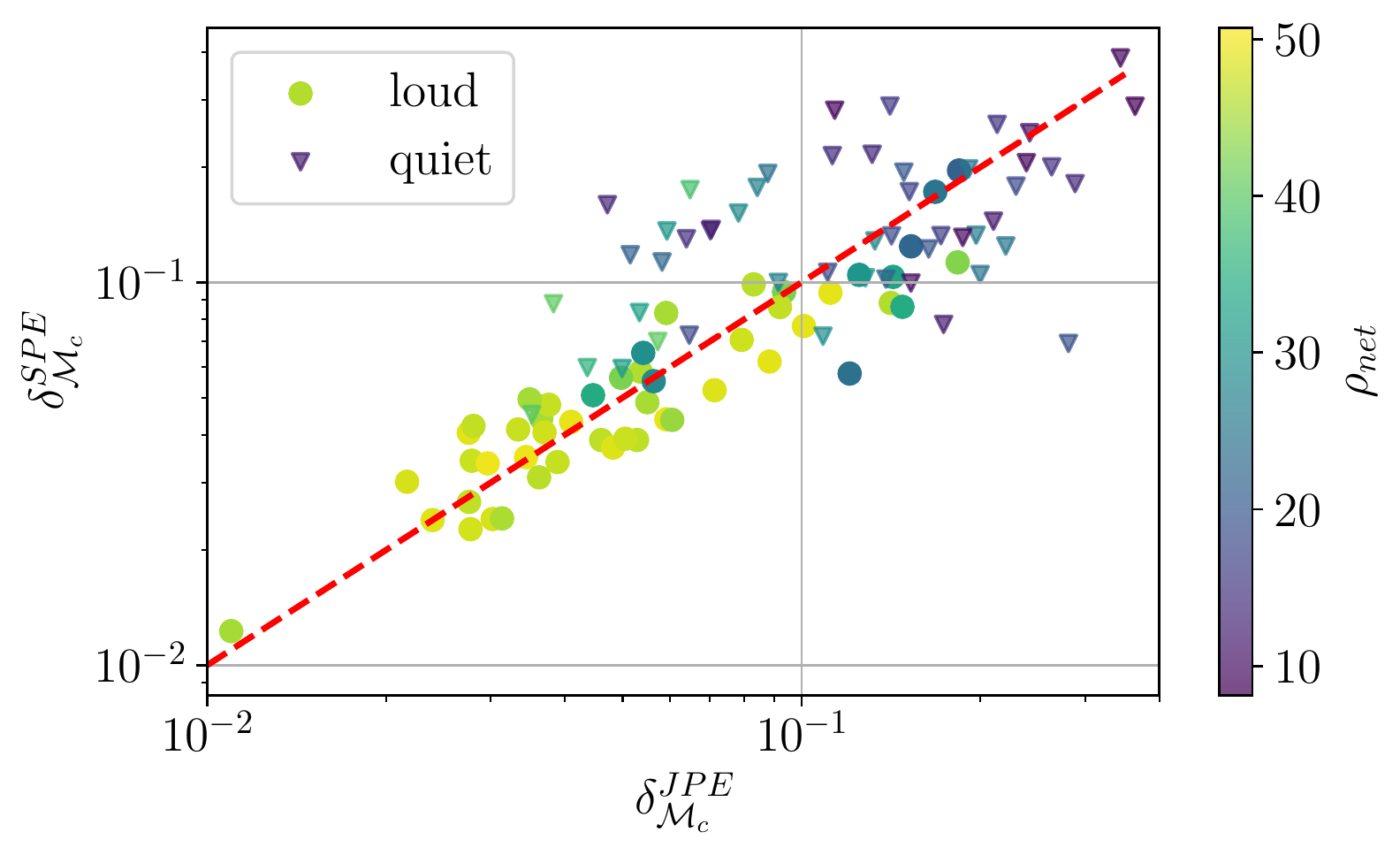}
    \caption{Comparison between the normalized width of the 90\% confidence interval for JPE and SPE in absence of noise. Since the spreads are very close for SPE and HS, the same relation is established for the JPE and HS comparison. There are larger differences between the two approaches but, globally, one is not better than the other as the events are approximately evenly distributed around the diagonal.}
    \label{fig:SpreadJPEnoNoise}
\end{figure}

Fig.~\ref{fig:SpreadJPEnoNoise} represents the normalized width of the posteriors for JPE versus SPE (which is comparable to JPE versus HS since HS versus SPE has widths aligning along the diagonal). There is more variance in this plot than for SPE vs HS. This is because the JPE method is significantly different from SPE, and we have a bigger variety of posteriors. Indeed, for JPE, we sometimes get broader posteriors but also tighter ones, depending on the characteristic of the two signals present in the data. However, the points are evenly distributed on the two sides of the diagonal, showing that, on average, none of the methods has widened posteriors compared to the other.

\end{document}